\documentclass[journal]{IEEEtran} 
\usepackage{cite} 

\usepackage[T1]{fontenc}    
\usepackage[utf8]{inputenc} 
\usepackage{newtxtext,newtxmath} 
\usepackage[english]{babel} 

\usepackage{amsmath, amsfonts, mathtools} 
\usepackage{physics} 

\DeclareMathOperator{\sinc}{sinc}

\usepackage{graphicx}
\usepackage[caption=false, font=footnotesize]{subfig} 
\usepackage{booktabs}  
\usepackage{array}     
\usepackage{tabularx}  
\newcolumntype{Y}{>{\centering\arraybackslash}X} 
\usepackage{booktabs}
\usepackage{caption}
\usepackage{float}
\usepackage{lscape}
\usepackage{multirow}
\usepackage{adjustbox}

\usepackage[ruled,vlined]{algorithm2e}  
\RestyleAlgo{ruled}  

\usepackage[hidelinks]{hyperref} 

\usepackage{textcomp}
\usepackage{verbatim}

\usepackage{stfloats} 

\usepackage{url} 

\addto\captionsenglish{}
\usepackage{balance}
\begin{document}
\title{Adaptive Channel Estimation and Quantized Feedback for RIS Assisted Optical Wireless Communication Systems}
\author{Muhammad  Khalil, \textit{Member, IEEE},  Ke Wang,  \textit{Senior, IEEE},  and Jinho Choi, \textit{Fellow, IEEE} 
  
\thanks{The authors Khalil and Wang are with the School of Engineering, RMIT University, Melbourne, Australia. Emails:$\,$ muhammad.khalil@rmit.edu.au,  and  ke.wang@rmit.edu.au. The author Choi with  the University of Adelaide, Adelaide, Australia. Email: jinho.choi@adelaide.edu.au}
}
\maketitle

\begin{abstract}

This paper develops a unified modeling, estimation, and feedback framework for reconfigurable intelligent surface (RIS)–assisted optical wireless links. The central modeling contribution is the introduction of a long-exposure pixel gain that extends the classical diffraction-limited response by statistically averaging angular jitter and mispointing. This formulation admits an exact real-integral representation and accurately reproduces two key physical effects: boresight attenuation and progressive sidelobe filling. Building upon this physically consistent model, the system integrates free-space path loss, Beer–Lambert atmospheric extinction, pixel-level diffraction, and optical efficiency with unitary-pilot least-squares (LS) channel estimation and quantized feedback. Analytical predictions closely agree with Monte Carlo simulations, providing concrete design guidelines. For instance, with a reconfigurable surface of \(N=64\) pixels, a pilot length of \(M=2N\), and a pilot signal-to-noise ratio (SNR) of \(20\,\mathrm{dB}\), the normalized mean-squared error (NMSE) is approximately \(0.005\), corresponding to a \(0.5\%\) effective-SNR reduction and a capacity penalty of only \(0.007\,\mathrm{bits/s/Hz}\). Phase feedback with six-bit quantization produces no perceptible degradation, establishing a practical benchmark for feedback resolution. Training overhead is shown to scale strongly with pixel geometry: halving the pixel width (and thus quartering the area) increases the required pilot length by roughly a factor of four.  Collectively, the framework reconciles physical-optics modeling with estimation-and-feedback design and provides a principled basis for scalable link budgeting in RIS-assisted optical networks.

\end{abstract}

\section{Introduction}
Optical wireless communication (OWC), including free-space optical (FSO) and visible-light communication (VLC), is a promising enabler of high-throughput, low-latency connectivity in beyond-5G networks~\cite{Fang2024}. By exploiting the unlicensed optical spectrum and compact light sources (lasers and LEDs), OWC systems routinely demonstrate multi-Gbps links in deployments ranging from indoor Li-Fi to inter-building backhaul and UAV downlinks~\cite{Qiu2022,Mahdy2004}. Nevertheless, OWC channels are intrinsically directional and line-of-sight (LoS) dependent; modest pointing errors or transient obstructions can precipitate severe alignment losses, while turbulence and adverse weather (e.g., fog, scintillation) introduce stochastic fading and attenuation~\cite{Krishnamoorthy2025,Zhu2002}. The ensuing penalties in outage and error floors have been documented repeatedly~\cite{Zhao2010}, motivating architectures that enhance robustness without eroding spectral efficiency.

Reconfigurable intelligent surfaces (RIS) have consequently attracted substantial interest as a means to engineer the propagation environment using programmable, largely passive reflecting elements~\cite{Khalil2025,Khalil2022}. First developed in the radio frequency (RF) domain~\cite{Wu2019a,Wu2020a} and now migrating to optics, RIS can redirect beams to recover blocked paths, concentrate energy into non-LoS regions, and elevate received power in coverage holes~\cite{Aboagye2024}. Empirical studies corroborate these benefits: mirror-array assistance can strengthen FSO link budgets and temper turbulence-induced fades~\cite{Naik2022}, while metasurface-aided VLC mitigates dead zones by steering or diffusing illumination~\cite{Abdelhady2021}. However, translating such demonstrations into scalable OWC systems is hindered by three intertwined factors that are particularly acute in the optical regime. 

First, the cascaded transmitter–RIS–receiver channel is high-dimensional: the number of effective coefficients scales with the RIS size, inflating pilot overhead and compressing the viable training window under turbulence or mobility~\cite{Ashok2019,Ardah2021,Abumarshoud2022,Haghshenas2024,Aboagye2023,Amirabadi2020}. With intensity-modulation/direct-detection (IM/DD), phase cannot be observed directly~\cite{Chaaban2022}, and millisecond-scale channel aging penalizes elongated pilots~\cite{Roa2025}. Second, feedback is intrinsically constrained: many OWC deployments are non-reciprocal and must convey channel state information (CSI) or RIS configurations over low-rate RF/IR control links~\cite{Aletri2019,Zhao2025}; unless carefully designed, coarse quantization of CSI or phase commands erodes coherent combining gains~\cite{Chen2023,Li2024,He2025}. Third,  hardware non-idealities limit ideal beam control: microelectromechanical systems (MEMS) mirror arrays offer agile steering but have limited phase agility and field-of-view~\cite{Sun2025,Wu2024,Herle2023}; optical metasurfaces approach $0$–$2\pi$ phase control but incur losses, discrete phase states, and millisecond switching~\cite{Zhang2023b,Abdollahramezani2022,Neuder2024,Wu2023}. Hence, a practical design must co-optimize training, feedback, and hardware constraints within a unified framework.
Many recent optical RIS studies assume ideal conditions: diffraction-limited apertures, perfect alignment, continuous-phase RIS elements, and error-free CSI feedback~\cite{Goodman2017,Born1999,Sun2025a,Shin2022}. However, in realistic OWC scenarios, pointing jitter weakens the boresight and fills sidelobe nulls, while non-reciprocal, rate-limited uplinks require compressed and quantized CSI. Existing models thus overlook how pixel-level optics under jitter affect pilot lengths, feedback resolution, and computational complexity under IM/DD constraints. Unlike~\cite{Sun2025a}, which assumes ideal diffraction and implicit feedback, we incorporate long-exposure (jitter-aware) optics and derive quantization-aware feedback rules that link pixel geometry and training overhead to Normalized Mean-Squared Error (NMSE) and capacity loss. Our unified framework embeds this physical model into an IM/DD-compatible unitary-pilot Least Squares (LS) estimator and a compressed-feedback mechanism for realistic uplinks, yielding closed-form prescriptions for pilot length and feedback resolution with favorable complexity and negligible performance loss.


\begin{table}[t]
\caption{RF vs Optical RIS (this work: bi-directional, non-reciprocal carriers)}
\label{tab:ris_rf_optical}
\centering
\footnotesize
\begingroup
\setlength{\tabcolsep}{4pt}
\renewcommand{\arraystretch}{1.1}
\begin{tabularx}{\columnwidth}{>{\bfseries}p{2.35cm} Y Y}
\toprule
Aspect & \centering\textbf{RF RIS} & \centering\textbf{Optical RIS (this work)} \tabularnewline
\midrule
Duplexing / channel symmetry &
TDD; \emph{reciprocal}. UL pilots calibrate DL~\cite{Wu2019a,Wu2020a}. &
\emph{Bi-directional, non-reciprocal}: optical DL + RF/IR UL; explicit CSI/phase feedback required. \\
\addlinespace[1pt]
Coherence time &
ms–hundreds ms (env.-dependent). &
Shorter, more variable; turbulence + sub-degree jitter~\cite{Zhu2002,Roa2025}. \\
\addlinespace[1pt]
Pilot / feedback overhead &
Lower via reciprocity; mature CSI compression~\cite{Wu2019a,Wu2020a}. &
Scales with $N$ (cascaded link); quantized UL feedback is the bottleneck~\cite{Ardah2021,Aboagye2023,Chen2023,Li2024}. \\
\bottomrule
\end{tabularx}
\endgroup
\end{table}

To substantiate this framework and address the identified research gap, this study provides the following key contributions:
\begin{itemize}
    \item \textbf{Jitter-aware long-exposure pixel model} with an exact real-integral formulation that predicts boresight attenuation and sidelobe-null smoothing under small angular jitter—effects not captured by ideal rectangular-aperture patterns~\cite{Goodman2017,Born1999}.
    \item \textbf{OWC-tailored estimation/feedback co-design}: a unitary-pilot LS estimator with \emph{processed, quantized} feedback that compresses the cascaded Tx–RIS–Rx estimate for bandwidth-limited RF/IR uplinks and aligns with IM/DD constraints~\cite{Aletri2019,Zhao2025}.
    \item \textbf{Closed-form sizing rules with complexity guarantees} relate pilot length, pixel geometry, and feedback resolution to a target NMSE with a negligible capacity penalty, thereby eliminating per-frame inversions and achieving near-quadratic complexity in array size.
   \item \textbf{Validated practicality:} Analytical predictions closely matched Monte-Carlo evaluations, confirming the robustness of the framework. In a representative setting (64-element RIS, moderate pilot SNR), the estimated NMSE remained well below \(10^{-2}\), implying negligible effective-SNR and capacity loss. Six-bit phase quantization delivered near-ideal performance, providing a practical benchmark for feedback resolution. Training effort scaled predictably with pixel geometry; smaller pixels required proportionally longer pilot sequences, while the feedback SNR reflected atmospheric extinction. Collectively, these findings affirm the scalability and physical consistency of the proposed optical RIS modeling and control strategy.
\end{itemize}

In contrast to prior RIS-OWC studies that decouple physical optics from system design (e.g.,~\cite{Sun2025a}), the present framework provides a direct, quantitative bridge from a jitter-aware pixel response and realistic optical efficiency to prescriptive pilot and feedback budgets under non-reciprocal control links.

The remainder of the paper is organized as follows. Section~\ref{sec:Sys_mod} introduces the RIS-assisted optical system model, which includes cascaded propagation and pixel-level optics. Section \ref{sec:Adapt } presents the proposed pilot-aided estimation and quantized feedback scheme, analyzing its dependence on training and quantization. Section \ref{sec: Feed } derives practical rules for balancing pilot length, feedback rate, and RIS design. Section \ref{sec:Simu}  validates the theory through Monte Carlo simulations. Section \ref{sec:Conclusion} concludes with future directions for scalable RIS-assisted OWC systems.

\section{\label{sec:Sys_mod}System model}

\begin{table}[t]
\caption{Main symbols and parameters used in the system model}
\label{tab:notation}
\centering
\footnotesize
\renewcommand{\arraystretch}{1.15}
\begin{tabular}{>{\raggedright\arraybackslash}p{1.9cm}%
                >{\raggedright\arraybackslash}p{4.8cm}%
                >{\centering\arraybackslash}p{0.9cm}}
\toprule
\textbf{Symbol} & \textbf{Physical Meaning / Definition} & \textbf{Units} \\
\midrule
$\mu^{(h)}_{x,n}, \mu^{(h)}_{y,n}$ &
Direction cosines of the $n$-th RIS pixel for hop $h\!\in\!\{\text{TR,RR}\}$ &
-- \\

$\Sigma^{(h)}$ &
Angular jitter covariance matrix for hop $h$ &
rad$^2$ \\

$\sigma^{(h)}_{x,n}, \sigma^{(h)}_{y,n}$ &
RMS pointing jitter in the $x$/$y$ directions &
mrad \\

$\rho^{(h)}_n$ &
Obliquity factor $\cos\psi_\mathrm{in}\cos\psi_\mathrm{out}$ &
-- \\

$S^{(h)}_n$ &
Strehl factor representing residual intra-aperture phase error &
-- \\

$A_n=\Delta x\Delta y$ &
Effective aperture area of the $n$-th RIS pixel &
m$^2$ \\

$\eta_{\mathrm{opt},n}$ &
Optical efficiency (reflectivity × polarization × insertion loss) &
-- \\

$k = 2\pi/\lambda$ &
Optical wavenumber corresponding to wavelength $\lambda$ &
rad/m \\

$G_{\mathrm{pix}}(\mu_x,\mu_y)$  & 
Pixel diffraction gain: ideal $G_0$ in \eqref{eq: Gpix}; long-exposure $\bar G_{\mathrm{pix}}$ in \eqref{eq:LE_def}, \eqref{eq:perhop_integral} &
-- \\

$\alpha$ &
Atmospheric extinction coefficient &
m$^{-1}$ \\

$d^{(\text{TR})}_n, d^{(\text{RR})}_n$ &
Propagation distances for Tx→RIS and RIS→Rx hops &
m \\

\bottomrule
\end{tabular}
\end{table}

\noindent
 
We consider an RIS-assisted  OWC system consisting of a transmitter ($\mathrm{\mathtt{T}}$), an RIS with $N$ individually controlled reflecting pixels (or elements), and a receiver ($\mathrm{\mathtt{R}}$), as illustrated in Fig.~\ref{fig:0}. 
The transmitter is located at $\mathbf{p}_{\mathrm{T}} = (0, 0, 0)$. 
The RIS is placed on the plane $z = d_{\mathrm{TR}}$, with the center of its $n^{\mathrm{th}}$ pixel located in the three-dimensional coordinate system at 
$\mathbf{p}_{n} = (x_{n}, y_{n}, d_{\mathrm{TR}})$, 
where $n = 1, \ldots, N$. 
The receiver aperture is positioned at 
$\mathbf{p}_{\mathrm{R}} = (x_{\mathrm{R}}, y_{\mathrm{R}}, z_{\mathrm{R}})$ 
where $z_{\mathrm{R}} > d_{\mathrm{TR}}$. 
The optical double-hop propagation distances are defined by the corresponding Euclidean norms of the transmitter–RIS and RIS–receiver vectors, as follows:  
\begin{figure}
\begin{centering}
\includegraphics[width=3.5in,viewport=2bp 0bp 550bp 350bp]{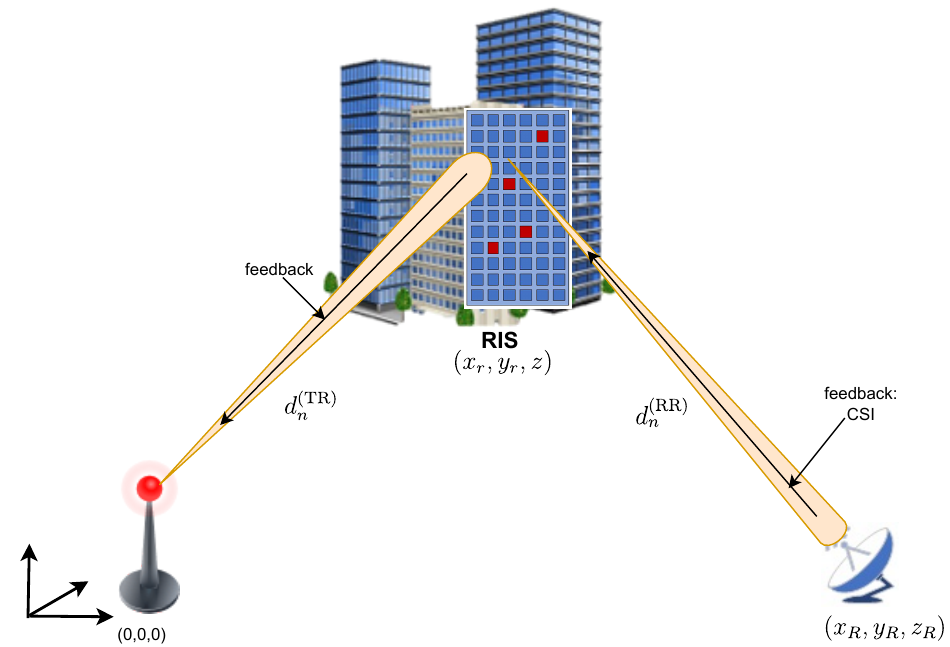}
\par\end{centering}
\caption{Geometric layout of the RIS-assisted OWC system, including transmitter, RIS, and receiver configurations \label{fig:0}}
\end{figure}

\begin{equation}
d_{n}^{(\mathrm{TR})}=\left\Vert \mathbf{p}_{n}-\mathbf{p}_{T}\right\Vert ,\quad d_{n}^{(\mathrm{RR})}=\left\Vert \mathbf{p}_{R}-\mathbf{p}_{n}\right\Vert ,\label{eq: dis}
\end{equation}
where $d_{n}^{(\mathrm{TR})}$  denotes the distance from the transmitter to the $n^{th}$ RIS pixel, and $d_{n}^{(\mathrm{RR})}$ denotes the distance from the $n^{th}$ RIS pixel to the receiver. Each RIS pixel is modeled as a uniformly illuminated rectangular aperture of size $\Delta x\times\Delta y$ (with an aperture area of $A_{n}\simeq\Delta x\,\Delta y$). Under the Fraunhofer (far-field) approximation, the normalized angular (intensity) gain of a single pixel is separable in the two transverse directions and equals the product of the squared $\mathrm{sinc}$ functions \cite{Goodman2017,Born1999}. In the analysis that follows, we adopt standard assumptions for OWC alignment, namely: (i) operation in the far-field (Fraunhofer) regime, such that angular diffraction by direction-cosines is valid; (ii) small-angle pointing jitter, with RMS $\sigma_{\alpha}\!\ll\!1$ rad (typically $\lesssim$\,5 mrad) and within the range where $k\,\sigma_{\alpha}\,\Delta\alpha \ll 1$ for $\alpha\!\in\!\{x,y\}$; and (iii) zero-mean Gaussian pointing errors with per-hop covariance $\Sigma^{(h)}$. These conditions justify adopting a long-exposure (jitter-averaged) formulation, beginning with the ideal diffraction response:
\begin{align}
G_{0}(\mu_{x}, \mu_{y})
&= \operatorname{sinc}^{2}(k_{x}\mu_{x})\,\operatorname{sinc}^{2}(k_{y}\mu_{y}),
\label{eq: Gpix}
\end{align}
where  $k_{x}=\tfrac{\pi \Delta x}{\lambda}$ and $k_{y}=\tfrac{\pi \Delta y}{\lambda}$ are the normalized spatial-frequency factors along $x$ and $y$; 
$\operatorname{sinc}(u)\triangleq \tfrac{\sin u}{u}$; 
$\mu_{x}\approx \sin\varphi_{x}$ and $\mu_{y}\approx \sin\varphi_{y}$ denote the direction cosines under the paraxial approximation; 
$\delta_{x}$ and $\delta_{y}$ are zero-mean angular jitters (radians) about the pixel normal; 
$S\simeq e^{-\sigma_{\phi}^{2}}$ is the Strehl factor (residual intra-aperture phase error); 
and $\rho\simeq \cos\psi^{\mathrm{in}}\cos\psi^{\mathrm{out}}$ is the obliquity (projection) factor.

In practice, the incident and reflected beams experience small angular perturbations due to residual tilt, tracking errors, pointing jitter, and atmospheric turbulence. 
We therefore adopt a long-exposure (statistical) model in which the instantaneous pointing angles fluctuate around the geometric look angles. 
With the angular jitters $\Delta_{x}$ and $\Delta_{y}$ as defined above, and the multiplicative corrections $S$ and $\rho$ accounting for residual phase error and obliquity, respectively, the long-exposure pixel response is
\begin{gather}
\bar{G}_{\mathrm{pix}}(\mu_{x},\mu_{y})=\nonumber \\
S\,\rho\;\mathbb{E}_{\delta_{x},\delta_{y}}\left[\mathrm{sinc}^{2}\big(k_{x}(\mu_{x}+\delta_{x})\big)\,\mathrm{sinc}^{2}\!\big(k_{y}(\mu_{y}+\delta_{y})\big)\right].\label{eq:LE_def}
\end{gather}
For the RIS-assisted link, \eqref{eq:LE_def} is evaluated for each hop 
$h \in \{\mathrm{TR}, \mathrm{RR}\}$, taking into account the respective geometric configurations and jitter statistics. 
Let $\boldsymbol{\mu}_{n}^{(h)} = [\mu_{x,n}^{(h)},\, \mu_{y,n}^{(h)}]^{\top}$ denote the geometric direction-cosine vector of pixel $n$ on hop $h$, 
and model the angular jitter vector as a correlated Gaussian random variable as follows:
We model small pointing errors on hop $h\!\in\!\{\mathrm{TR},\mathrm{RR}\}$ as zero-mean Gaussian jitter:
\begin{equation}
\boldsymbol{\delta}^{(h)} =
\begin{bmatrix}
\delta_x^{(h)}\\[2pt]
\delta_y^{(h)}
\end{bmatrix}
\sim \mathcal{N}\!\big(\mathbf{0},\,\boldsymbol{\Sigma}^{(h)}\big),\qquad
\boldsymbol{\Sigma}^{(h)} \succeq \mathbf{0}.
\label{eq:jitter_model}
\end{equation}
with a common per-hop covariance for notational economy (pixel-specific covariances are recovered by $\boldsymbol{\Sigma}^{(h)}\!\to\!\boldsymbol{\Sigma}^{(h)}_n$ where needed).

Given this jitter, the long-exposure per-hop pixel gain for element $n$ is
\begin{gather}
\bar{G}_{\mathrm{pix},n}^{(h)}=\nonumber \\
S_{n}^{(h)}\,\rho_{n}^{(h)}\;\mathbb{E}_{\boldsymbol{\delta}^{(h)}}\!\Big[\mathrm{sinc^{2}}\!\big(k_{x}(\mu_{x,n}^{(h)}+\delta_{x}^{(h)})\big)\,\mathrm{sinc^{2}}\!\big(k_{y}(\mu_{y,n}^{(h)}+\delta_{y}^{(h)})\big)\Big].\label{eq:perhop_LE}
\end{gather}

Evaluating \eqref{eq:perhop_LE} in the Fourier domain yields the compact real form
\begin{multline}
\bar{G}_{\mathrm{pix},n}^{(h)}
=\frac{S_{n}^{(h)}\,\rho_{n}^{(h)}}{4k_{x}k_{y}}
\int_{-2k_{x}}^{2k_{x}}\!\!\int_{-2k_{y}}^{2k_{y}}
\Big(1-\tfrac{|\omega_{x}|}{2k_{x}}\Big)\,
\Big(1-\tfrac{|\omega_{y}|}{2k_{y}}\Big)\\
\exp\!\Big(-\tfrac{1}{2}\,\boldsymbol{\omega}^{\!\top}\boldsymbol{\Sigma}^{(h)}\boldsymbol{\omega}\Big)\,
\cos\!\big(\boldsymbol{\omega}^{\!\top}\boldsymbol{\mu}_{n}^{(h)}\big)\,
d\omega_{y}\,d\omega_{x},\quad
\boldsymbol{\omega}=\begin{bmatrix}\omega_{x}\\ \omega_{y}\end{bmatrix}.
\label{eq:perhop_integral}
\end{multline}

 The triangular windows arise from the Fourier transform of $\sinc^2$; the Gaussian factor represents the jitter characteristic function. Increasing $\boldsymbol{\Sigma}^{(h)}$ broadens the main lobe, lowers the boresight peak, and fills the nulls; anisotropy and correlation follow from the diagonal/off-diagonal entries of $\boldsymbol{\Sigma}^{(h)}$.

The two-hop long-exposure pixel gain factors as the product of the hops:
\begin{gather}
\bar{G}_{\mathrm{pix},n}=\prod_{h\in\{\mathrm{TR},\mathrm{RR}\}}\bar{G}_{\mathrm{pix},n}^{(h)},\qquad\nonumber \\
0\le\bar{G}_{\mathrm{pix},n}\le\big(S_{n}^{(\mathrm{TR})}\rho_{n}^{(\mathrm{TR})}\big)\big(S_{n}^{(\mathrm{RR})}\rho_{n}^{(\mathrm{RR})}\big),\label{eq:twohop}
\end{gather}
which, upon substituting $\eqref{eq:perhop_integral}$, yields the final expression used in the Appendix. For completeness, the independent-jitter reduction proceeds by folding the symmetric integration domain to $[0,2k_{x}]\times[0,2k_{y}]$, after which the double integral factors into the product of two one-dimensional blur kernels $\mathcal{B}(\cdot;k_{\alpha},\sigma_{\alpha})$.
The aperture area $A_{n}$ controls the on-axis gain scaling, consistent with the aperture-gain relation $G_{max}\propto A_{n}/\lambda^{2}$
\cite{Balanis2016}.

To ensure consistent optical power normalization, we distinguish between static, hardware-dependent efficiency and dynamic, software-defined RIS control. The static optical efficiency of the $n$-th pixel is defined as
\begin{equation}
\eta_{\mathrm{opt},n}\;\triangleq\;R_{n}(\lambda)\,\xi_{p,n}\,L_{\mathrm{ins},n},\label{eq: Eq8}
\end{equation}
 where \(R_{n}(\lambda)\) is the wavelength–dependent reflectivity, \(\xi_{p,n}\) is the polarization efficiency, and \(L_{\mathrm{ins},n}\in(0,1]\) accounts for insertion losses due to the substrate and metal/dielectric layers. Thus, in \eqref{eq: Eq8} and throughout the model, \(\eta_{\mathrm{opt},n}\) \emph{already subsumes} reflectivity, polarization, and insertion losses and is applied \emph{once} per pixel; these factors are not re-multiplied elsewhere. When field amplitudes are used, the corresponding scaling is \(\sqrt{\eta_{\mathrm{opt},n}}\) (power–domain factors use \(\eta_{\mathrm{opt},n}\)). By contrast, geometric/propagation terms—e.g., obliquity \(\rho^{(h)}_n\), Strehl factor \(S^{(h)}_n\), pixel area \(A_n\), long-exposure diffraction gain \(G_{\mathrm{pix}}(\cdot)\), and atmospheric extinction—are modeled explicitly outside \(\eta_{\mathrm{opt},n}\). The programmable RIS control operates through phase states (and, if available, coarse amplitude states) and is independent of static efficiency.

In contrast, the programmable control applied at runtime is expressed as
\begin{equation}
\theta_{n}=\;\rho_{n}\,e^{j\phi_{n}},
\end{equation}
 where $\rho_{n}\left(0<\rho_{n}\le1\right)$ is the programmable amplitude scaling, which is typically set to 1 for a pure-phase RIS. Importantly, $\rho_{n}$ appears only inside $\theta_{n}$ and is excluded from $\eta_{\mathrm{opt},n}.$ This prevents double counting of attenuation and ensures that hardware characterization (through $\eta_{\mathrm{opt},n}$)
remains independent of algorithmic optimization (through $\theta_{n}$)

The line-of-sight (LOS) optical field contributions of pixel $n$ are modeled using the transmitter and receiver directivities $G_{T}$ and $G_{R}$, the Beer–Lambert extinction coefficient $\alpha$, and the optical wavenumber $k=2\pi/\lambda$. The complex LOS field gains for the two hops ($\mathtt{T}\!\to\!n$ and $n\!\to\!\mathtt{R}$) are
\begin{gather}
h_{n}^{(\mathrm{TR})}=\sqrt{\frac{A_{n}\,G_{T}\,\bar{G}_{\mathrm{pix},n}^{(\mathrm{TR})}\,\eta_{\mathrm{opt},n}}{(4\pi d_{n}^{(\mathrm{TR})})^{2}}}\;e^{-\frac{\alpha}{2}d_{n}^{(\mathrm{TR})}}\;e^{-jkd_{n}^{(\mathrm{TR})}}\;\zeta_{n}^{(\mathrm{TR})},\nonumber \\[4pt]
h_{n}^{(\mathrm{RR})}=\sqrt{\frac{A_{n}\,G_{R}\,\bar{G}_{\mathrm{pix},n}^{(\mathrm{RR})}\,\eta_{\mathrm{opt},n}}{(4\pi d_{n}^{(\mathrm{RR})})^{2}}}\;e^{-\frac{\alpha}{2}d_{n}^{(\mathrm{RR})}}\;e^{-jkd_{n}^{(\mathrm{RR})}}\;\zeta_{n}^{(\mathrm{RR})},\label{eq: h=000026h}
\end{gather}
where $\zeta_{n}^{(\cdot)}\!\sim\!\mathcal{CN}(0,1)$ models small-signal turbulence. The cascaded per-element coefficient (pre-control) is
\begin{equation}
g_{n}\;\triangleq\;h_{n}^{(\mathrm{RR})}\,h_{n}^{(\mathrm{TR})}, 
\qquad 
\mathbf{g}\;\triangleq\;[\,g_{1},\ldots,g_{N}\,]^{\mathsf T}.
\label{eq: gn}
\end{equation}

Under the paraxial approximation, $\mu_{\alpha,n}^{(h)}=\sin\!\big(\theta_{\alpha,n}^{(h)}\big)\approx \theta_{\alpha,n}^{(h)}$ (radians) is applicable for $\alpha\in\{x,y\}$ and $h\in\{\mathrm{TR},\mathrm{RR}\}$.

We also collect the RIS weights in $\boldsymbol{\theta}=[\theta_{1},\ldots,\theta_{N}]^{\mathsf{T}},$
using the diagonal control matrix $\mathrm{\Theta=diag(\boldsymbol{\theta})}$. 

The coefficient $g_{n}$ in \eqref{eq: gn} is determined by geometry, free-space loss, pixel diffraction, and optical efficiency. 
To capture realistic effects, each hop coefficient is expressed as a deterministic baseline multiplied by a random irradiance factor, as follows:
\begin{equation}
h_{n}^{TR}\;\longrightarrow\;h_{n}^{TR}\,\sqrt{H_{n}^{TR}},\qquad h_{n}^{RR}\;\longrightarrow\;h_{n}^{RR}\,\sqrt{H_{n}^{RR}},
\end{equation}
where $H_{n}^{TR}$ and $H_{n}^{RR}$ are positive random variables that capture atmospheric scintillation, which is log-normal under weak turbulence and Gamma-Gamma under moderate-to-strong turbulence. Consequently, the cascaded coefficient becomes
\begin{equation}
g_{n}\longrightarrow\underbrace{h_{n}^{RR}h_{n}^{TR}}_{\underset{\textrm{baseline}}{\text{deterministic}}}.\underbrace{\sqrt{H_{n}^{TR}H_{n}^{RR}}}_{\text{turbulence factor}},\label{eq: ga}
\end{equation}
and the instantaneous cascaded power gain is
\begin{equation}
|g_{n}|^{2}\;=\;\big|h_{n}^{RR}h_{n}^{TR}\big|^{2}\,H_{n}^{RR}H_{n}^{TR}.\label{eq: gn^2}
\end{equation}

The deterministic baseline term $\big|h_{n}^{\mathrm{RR}}h_{n}^{\mathrm{TR}}\big|^{2}$ accounts for aperture size $A_{n}$, free-space path loss, directivities $G_{T}, G_{R}$, and pixel diffraction $G_{\mathrm{pix}}$. With turbulence normalized such that $\mathbb{E}[H_{n}^{\mathrm{TR}}]=\mathbb{E}[H_{n}^{\mathrm{RR}}]=1$, the mean cascaded per-element power is
\begin{gather}
\mathbb{E}\!\left[|g_{n}|^{2}\right]\;\approx\;\,\eta_{\mathrm{opt},n}^{2}\,e^{-2\alpha\,(d_{n}^{TR}+d_{n}^{RR})}\nonumber \\
\frac{A_{n}^{2}G_{T}G_{R}\,G_{\mathrm{pix}}(\mu_{x,n}^{TR},\mu_{y,n}^{TR})\,G_{\mathrm{pix}}(\mu_{x,n}^{RR},\mu_{y,n}^{RR})}{(4\pi)^{4}\,\left(d_{n}^{TR}\right)^{2}\left(\,d_{n}^{RR}\right)^{2}}.\label{eq: E=gn_abs}
\end{gather}

The received baseband signal in the data phase has finally become 
\begin{equation}
y_{f}\;=\;\sqrt{P_{d}}\;\mathbf{g}^{H}\boldsymbol{\theta}\,x\;+\;n_{f},
\end{equation}
where $P_{d}$ is the transmitted data power, $\mathbf{g}^{H}$
is the Hermitian (conjugate transpose) of the $N\times1$ cascaded channel vector $\boldsymbol{g}$, $x$ is a complex modulation symbol with unit average energy, i.e., $\mathbb{E}\mid x\mid^{2}=1,$, and $n_{f}\sim\mathcal{CN}(0,\sigma_{\mathrm{tot}}^{2})$ is zero-mean additive receiver noise with total variance 
\begin{equation}
\sigma_{\mathrm{tot}}^{2}=\sigma_{\mathrm{shot}}^{2}+\sigma_{\mathrm{th}}^{2},\label{eq: temp}
\end{equation}

\noindent measured over the receiver\textquoteright s equivalent noise bandwidth $B$. 

The shot-noise variance (current domain) \cite{Khalighi2014} is modeled as 
\begin{equation}
\sigma_{\mathrm{shot}}^{2}\;=\;2q\left(I_{\mathrm{sig}}+I_{\mathrm{bg}}+I_{\mathrm{dark}}\right)B,\label{eq: temp_sh}
\end{equation}
 where $q$ is the electron charge, $I_{\mathrm{sig}}=\mathcal{R}P_{\mathrm{sig}}$ is the average photocurrent due to the useful received optical power $P_{\mathrm{sig}}$, $I_{\mathrm{bg}}=\mathcal{R}P_{\mathrm{bg}}$ is the background-light photocurrent due to ambient optical power
$P_{\mathrm{bg}}$, and $I_{\mathrm{dark}}$ is the contribution of the dark current. Here, $\mathcal{R}$ denotes the photodiode responsivity (A/W) \cite{Kahn1997}. 

The thermal-noise variance (current domain), due to the transimpedance front-end of the receiver, can be expressed as
\begin{align}
\sigma_{\mathrm{th}}^{2}\;= & \;\frac{4k_{B}T}{R_{F}}\,I_{2}R_{b}\;+\;\frac{16\pi k_{B}T}{g_{m}R_{F}}\Bigg(\varGamma+\frac{1}{g_{m}R_{C}}\Bigg)C_{T}^{2}\,I_{3}R_{b}^{3}\nonumber \\
 & \;+\;\frac{4\pi^{2}k_{B}T}{g_{m}^{2}}\,C_{T}^{2}\,I_{f}R_{b}^{2},
\end{align}
where $k_{B}$ is Boltzmann's constant, $T$ is the receiver noise temperature, $R_{F}$ is the TIA feedback resistance, $g_{m}$ is the input transistor transconductance, $\varGamma$ is the device channel-noise factor, $R_{C}$ is a small series resistance, and $C_{T}$ is the total input capacitance (photodiode plus parasitics), and factors $I_{2}$, $I_{3}$, and $I_{f}$ are dimensionless noise-bandwidth
factors determined by the overall filter/equalizer defined by \cite{Kahn1997}.

\subsection{Expected SNR}
For a purely phase-shifting RIS ($\rho_{n}=1,\forall n$) with optimal phase control $\theta_{n}=e^{-j\arg(g_{n})}$, the received SNR after coherent combining is determined by the composite channel gains $g_{n}$.  
Since $|g_{n}|$ is random under turbulence, we consider its statistical expectation.  
The expected post-combining SNR is
\begin{equation}
\mathbb{E}[\gamma_{f}^{\star}]
=\frac{P_{d}}{\sigma_{\mathrm{tot}}^{2}}
\!\left(
\sum_{n=1}^{N}\mathbb{E}[|g_{n}|^{2}]
+\!\!\sum_{\substack{n,m=1\\ n\neq m}}^{N}
\mathbb{E}[|g_{n}|]\,
\mathbb{E}[|g_{m}|]
\right)\!,
\label{eq:expSNR}
\end{equation}
where $\mathbb{E}[|g_{n}|^{2}]$ represents the average per-element power determined by the deterministic geometry, path loss, and optical efficiency, while $\mathbb{E}[|g_{n}|]$ denotes the mean field amplitude under turbulence.   The latter depends on the turbulence regime, following the log-normal model under weak turbulence and the Gamma–Gamma distribution under moderate-to-strong conditions.  Turbulence normalization preserves the mean power, but the coherent combining gain is reduced according to the statistics of $\mathbb{E}[|g_{n}|]$.

In the log-normal regime (weak turbulence), let the log-intensity be normalized as $\ln H^{(\cdot)}\sim\mathcal{N}\!\big(-\tfrac{1}{2}\sigma_{\ln I,(\cdot)}^{2},\,\sigma_{\ln I,(\cdot)}^{2}\big)$ so that $\mathbb{E}[H^{(\cdot)}]=1$. Then $\mathbb{E}[\sqrt{H^{(\cdot)}}]=\exp\!\big(-\sigma_{\ln I,(\cdot)}^{2}/8\big)$
\cite{Sharma2020}, and by independence across the two hops, the average field amplitude per element satisfies
\begin{equation}
\mathbb{E}[\,|g_{n}|\,]=\sqrt{\mathbb{E}[\,|g_{n}|^{2}\,]}\;\exp\!\Big(-\tfrac{1}{8}\big(\sigma_{\ln I,\mathrm{TR}}^{2}+\sigma_{\ln I,\mathrm{RR}}^{2}\big)\Big).\label{eq: Eg2}
\end{equation}

Equivalently, if one prefers the log-amplitude parameterization $H^{(\cdot)}=\exp(2\chi^{(\cdot)})$
with $\chi^{(\cdot)}\sim\mathcal{N}\!\big(-\sigma_{\chi,(\cdot)}^{2},\,\sigma_{\chi,(\cdot)}^{2}\big)$ (which also enforces $\mathbb{E}[H^{(\cdot)}]=1$), then $\mathbb{E}[\sqrt{H^{(\cdot)}}]=\exp\!\big(-\sigma_{\chi,(\cdot)}^{2}/2\big)$
and $\sigma_{\chi,(\cdot)}^{2}=\sigma_{\ln I,(\cdot)}^{2}/4$ yield the same expression for $\mathbb{E}[\,|g_{n}|\,]$.

Under stronger turbulence, irradiance is well modeled by the mean-normalized Gamma-Gamma law \cite{Chatzidiamantis2011} with shape parameters $(\alpha,\beta)$ and density
\begin{equation}
f_{H}(h)=\frac{2(\alpha\beta)^{\frac{\alpha+\beta}{2}}}{\Gamma(\alpha)\Gamma(\beta)}\,h^{\frac{\alpha+\beta}{2}-1}\,K_{\alpha-\beta}\!\left(2\sqrt{\alpha\beta\,h}\right),\qquad h>0,
\end{equation}
where $K_{\nu}(\cdot)$ is the modified Bessel function of the second kind and $\mathbb{E}[H]=1$. In this case,
\begin{equation}
\mathbb{E}\!\big[\sqrt{H}\big]=\frac{\Gamma(\alpha+\tfrac{1}{2})\,\Gamma(\beta+\tfrac{1}{2})}{\Gamma(\alpha)\,\Gamma(\beta)\,\sqrt{\alpha\beta}},
\end{equation}
so, for independent hops with parameters $(\alpha_{\mathrm{TR}},\beta_{\mathrm{TR}})$ and $(\alpha_{\mathrm{RR}},\beta_{\mathrm{RR}})$,
\begin{equation}
\mathbb{E}[\,|g_{n}|\,]=\sqrt{\mathbb{E}[\,|g_{n}|^{2}]}\prod_{s\in\{\mathrm{TR},\mathrm{RR}\}}\frac{\Gamma(\alpha_{s}+\tfrac{1}{2})\,\Gamma(\beta_{s}+\tfrac{1}{2})}{\Gamma(\alpha_{s})\,\Gamma(\beta_{s})\,\sqrt{\alpha_{s}\beta_{s}}}\label{eq: Gam_gn}
\end{equation}

These relations make it explicit that, with $\mathbb{E}[H]=1$, the mean per-element power $\mathbb{E}[\,|g_{n}|^{2}\,]$ equals the deterministic
baseline, whereas the coherent combining term in $\mathbb{E}[\gamma_{f}^{\star}]$ is reduced according to the turbulence-dependent factor multiplying $\sqrt{\mathbb{E}[\,|g_{n}|^{2}\,]}$.

\subsection{Pilot-phase SNR}
During the pilot (training) phase, let the transmitter use pilot power $P_{T}$, and the RIS should apply a non-coherent or orthogonal training pattern $\mathbf{u}$ with $|u_{n}|=1$ and zero-mean cross-products across elements (e.g., random phases or unitary/orthogonal sequences). The received pilot sample is
\begin{equation}
y_{\mathrm{pilot}}=\sqrt{P_{T}}\;\sum_{n=1}^{N}g_{n}\,u_{n}+n_{\mathrm{pilot}},\qquad n_{\mathrm{pilot}}\sim\mathcal{CN}(0,\sigma_{\mathrm{tot}}^{2}),
\end{equation}
where the summation represents the aggregate pilot contribution from all RIS elements, and $\sigma_{\mathrm{tot}}^{2}$ denotes the total noise power in the pilot band. Under orthogonal or statistically independent training, all cross-terms vanish in expectation, i.e., $\mathbb{E}[g_{n}g_{m}^{*}u_{n}u_{m}^{*}]=0$ for $n\neq m$. Accordingly, the average pilot SNR simplifies to
\begin{equation}
\gamma_{\mathrm{pilot}}=\frac{P_{T}}{\sigma_{\mathrm{tot}}^{2}}\;\sum_{n=1}^{N}\mathbb{E}\!\left[|g_{n}|^{2}\right].\label{eq: Ypilot}
\end{equation}
where $\mathbb{E}[|g_{n}|^{2}]$ captures the mean per-element power determined by deterministic geometry, path loss, optical efficiency, and long-exposure diffraction gain. 
\noindent
Physically, $\gamma_{\mathrm{pilot}}$ depends on several optical and geometric parameters defined in Section~\ref{sec:Sys_mod}. In particular, it scales with the pixel area $A_{n}$ and optical efficiency $\eta_{\mathrm{opt},n}$, while decreasing with larger propagation distances between the transmitter, RIS, and receiver ($d_{\mathrm{TR}}$, $d_{\mathrm{RR}}$) and with stronger atmospheric extinction. The wavelength $\lambda$ also influences the SNR through diffraction and path-loss effects. These relationships link the physical RIS design parameters directly to the estimation performance, showing that $\gamma_{\mathrm{pilot}}$ bridges the optical configuration and the achievable training accuracy.

By contrast, if the RIS coherently aligns pilots with the instantaneous phases (i.e., $u_{n}=e^{-j\arg(g_{n})}$ during training), the expected SNR includes the coherent cross-term and becomes
\begin{equation}
\mathbb{E}\!\big[\gamma_{\mathrm{pilot}}^{\mathrm{(coh)}}\big]=\frac{P_{T}}{\sigma_{\mathrm{tot}}^{2}}\Biggl(\sum_{n=1}^{N}\mathbb{E}[|g_{n}|^{2}]+\sum_{\substack{n,m=1\\
n\neq m
}
}^{N}\mathbb{E}[|g_{n}|]\;\mathbb{E}[|g_{m}|]\Biggr),\label{eq: Y_coh}
\end{equation}

Here, $\mathbb{E}[|g_{n}|]$ depends on the turbulence regime: log-normal under weak turbulence and Gamma-Gamma under moderate to strong turbulence. Equation \eqref{eq: Ypilot} thus characterizes non-coherent accumulation, whereas \eqref{eq: Y_coh} quantifies the additional coherent gain achievable when the pilot phases are dynamically aligned.

\section{\label{sec:Adapt }Adaptive RIS Feedback Scheme}

Accurate and efficient channel state acquisition is essential for RIS phase control, particularly under time-varying conditions such as turbulence and mobility. We adopt a pilot-aided scheme: the transmitter sends $M$ pilots while the RIS cycles through $M$ known patterns collected in a column-orthonormal (semi-unitary) matrix $\boldsymbol{\Phi}\!\in\!\mathbb{C}^{M\times N}$ with $\boldsymbol{\Phi}^H\boldsymbol{\Phi}=M\mathbf I_N$ ($M\!\ge\!N$). The received pilot vector is
\[
\mathbf y=\sqrt{P_T}\,\boldsymbol{\Phi}\mathbf g+\mathbf n,\qquad 
\mathbf n\sim\mathcal{CN}(\mathbf0,\sigma^2\mathbf I_M).
\]
The LS estimate is 
\[
\hat{\mathbf g}=\frac{1}{M\sqrt{P_T}}\,\boldsymbol{\Phi}^H\mathbf y.
\]
With $\gamma_{\text{pilot}}\!\triangleq\!\frac{P_T\|\mathbf g\|^2}{\sigma^2}$, the normalized mean squared error (MSE) is
\[
\mathrm{NMSE}=\frac{N}{M\,\gamma_{\text{pilot}}}.
\]

After the vector of pilot measurements $\mathbf{y}$ has been collected, the receiver initiates the feedback process to deliver essential channel information to the transmitter, which operates entirely through the optical RIS elements in our architecture. There are two principal feedback strategies:

$\textbf{1) Raw (Direct) Feedback:}$ In the most basic method, the
receiver directly quantizes the measurement vector $\mathbf{y}$ and
transmits these quantized values over the RIS-assisted uplink. Letting $\mathcal{Q}_{y}(\cdot)$
denote scalar quantization applied elementwise to the pilot response vector$\mathbf{y}\in\mathbb{C}^{M}$, the actual feedback is $\mathbf{y}_{\mathrm{FB}}=\mathcal{Q}_{y}(\mathbf{y}).$ While this approach requires minimal processing at the receiver, it is rarely efficient: the dimension $M$ of $\mathbf{y}$ (number of pilots and RIS patterns) can be large, and transmitting all these quantized pilot responses over the optical RIS feedback link, which may have limited SNR and bandwidth, is typically infeasible in practice.

\textbf{2) Processed (Compressed) Feedback:} 
A more efficient strategy is for the receiver to locally estimate the cascaded RIS channel using a least–squares (LS) estimator based on the known pilot matrix $\boldsymbol{\Phi}$ and the stacked pilot observations $\mathbf{y}$,  which is given by
\begin{equation}
\hat{\mathbf{g}}_{\mathrm{Rx}}=(\boldsymbol{\Phi}^{H}\boldsymbol{\Phi})^{-1}\boldsymbol{\Phi}^{H}\mathbf{y}=\frac{1}{M}\,\boldsymbol{\Phi}^{H}\mathbf{y}.
\end{equation}
The estimated channel is quantized elementwise for feedback, i.e., \(\hat{\mathbf{g}}_{\mathrm{FB}}=\mathcal{Q}_{g}(\hat{\mathbf{g}}_{\mathrm{Rx}})\), where \(\mathcal{Q}_{g}\) is a \(b_g\)-bit quantizer applied to each coefficient of \(\hat{\mathbf{g}}_{\mathrm{Rx}}\in\mathbb{C}^{N}\). Compared with forwarding quantized pilots \(\mathbf{y}_{\mathrm{FB}}=\mathcal{Q}_{y}(\mathbf{y})\) of dimension \(M\), the processed scheme aligns the payload with unknowns rather than with pilots: \(B_{\mathrm{raw}}=M\,b_{y}\) versus \(B_{\mathrm{proc}}=N\,b_{g}\) (or \(N\,b_{\theta}\) for phase-only), with \(N\ll M\) in typical RIS training. Quantizing after estimation isolates feedback distortion from measurement noise, so performance scales cleanly with pilot length and bit depth. With unitary (orthogonal) training, LS is unbiased and separable, requiring only \(\mathcal{O}(MN)\) operations (no matrix inversion); in practice, the uplink rate, not estimator complexity, is the bottleneck.

To further reduce payload, the receiver can exploit sparsity and temporal correlation via compressed feedback, computing \(\hat{\mathbf{g}}_{\mathrm{FB}}=\mathcal{C}(\mathbf{W}^{H}\hat{\mathbf{g}}_{\mathrm{Rx}})\) with a sparsifying basis \(\mathbf{W}\) and a compressed (quantization–encoding) mapping \(\mathcal{C}(\cdot)\). Retaining only \(K\ll N\) dominant coefficients results in an accuracy improvement that is approximately \(\mathrm{NMSE}\propto 1/K\) (empirically, Fig.~12 is nearly inverse in \(K/N\)). At the transmitter, the decoded \(\hat{\mathbf{g}}_{\mathrm{FB}}\) is used directly to set the RIS phases for the next frame, completing the pilot \(\rightarrow\) local estimation \(\rightarrow\) (compressed/quantized) feedback \(\rightarrow\) RIS re-phasing loop that maintains robust performance under rate-limited uplinks and finite-resolution hardware.

The performance of this adaptive process depends directly on the accuracy
of the estimated channel vector $\hat{\mathbf{g}}_{\mathrm{FB}}$ that is fed back to the transmitter. This estimation accuracy is fundamentally determined by the Fisher information matrix,
\begin{equation}
\mathbf{J}=\frac{P_{T}}{\sigma^{2}}\boldsymbol{\Phi}^{H}\boldsymbol{\Phi}=\frac{P_{T}}{\sigma^{2}}\,M\mathbf{I}_{N},\;\mathrm{Cov}(\hat{\mathbf{g}})\succeq\mathbf{J}^{-1}=\frac{\sigma^{2}}{P_{T}M}\,\mathbf{I}_{N},\label{eq: J}
\end{equation}
which measures how much information about the channel is provided by the set of pilot observations, taking into account the noise variance $\sigma^{2}$.
According to the Cramér–Rao lower bound, the minimum achievable error covariance for any unbiased channel estimator equals the inverse of this matrix:
\begin{equation}
\mathrm{Cov}(\hat{\mathbf{g}})\succeq\mathbf{J}^{-1}=\sigma^{2}(\boldsymbol{\Phi}^{H}\boldsymbol{\Phi})^{-1}.
\end{equation}

To quantitatively assess the quality of the channel estimate, we use the normalized mean squared error (NMSE), defined as
\begin{gather}
\mathrm{NMSE}=\frac{\mathbb{E}\|\hat{\mathbf{g}}-\mathbf{g}\|^{2}}{\|\mathbf{g}\|^{2}}=\frac{\sigma^{2}}{P_{T}\|\mathbf{g}\|^{2}}\,\mathrm{tr}\!\left[(\boldsymbol{\Phi}^{H}\boldsymbol{\Phi})^{-1}\right]\nonumber \\
=\frac{N}{M\,\gamma_{\mathrm{pilot}}},\quad\gamma_{\mathrm{pilot}}\triangleq\frac{P_{T}\|\mathbf{g}\|^{2}}{\sigma^{2}},\label{eq: NMSE 14}
\end{gather}
where $\mathbf{g}$ denotes the true cascaded channel vector and $\hat{\mathbf{g}}$ its estimated counterpart.

A lower NMSE indicates higher estimation accuracy, which, in turn, leads to better RIS phase optimization and improved overall system performance. For a given system geometry and RIS design, and assuming additive white Gaussian noise (AWGN), the per-pilot SNR $\gamma_{\mathrm{pilot}}$ is determined by the cascaded channel gain, as defined in Section \ref{sec:Sys_mod}. Under LS estimation with an orthogonal (unitary) pilot matrix, the normalized NMSE satisfies
\begin{equation}
\textrm{NMSE}=\frac{N}{M\:\gamma_{\mathrm{pilot}}}.\label{eq: NMSE}
\end{equation}

Consequently, to achieve a target NMSE $\epsilon$, the pilot length must be satisfied:
\begin{equation}
M \;\ge\; \frac{N}{\,\epsilon\,\gamma_{\mathrm{pilot}}(\mathcal{D})\,},
\label{eq:M_req}
\end{equation}
where $\epsilon$ denotes the target normalized NMSE, $\gamma_{\mathrm{pilot}}$ is the pilot SNR (in linear scale), and $\mathcal{D}$ represents the complete set of RIS optical design parameters, 
\[
\mathcal{D}=\{\lambda,\,A_n,\,R(\lambda),\,G_{\mathrm{pix}},\,\rho_n,\,d_n^{(\mathrm{TR})},\,d_n^{(\mathrm{RR})}\}.
\]
Equation~\eqref{eq:M_req} explicitly links estimation accuracy, pilot overhead, and hardware configuration: stronger per-element optical gain or higher pilot SNR directly reduces the training length required to satisfy a given $\epsilon$. For illustration, at $\gamma_{\mathrm{pilot}}=20$\,dB with $N=64$ RIS elements and a target NMSE of $\epsilon=0.01$, approximately $M\!\approx\!128$ pilot symbols are required. This numerical example provides an intuitive view of how pilot length, estimation fidelity, and the optical design variables summarized in $\mathcal{D}$ jointly shape the system’s training requirements.

Once a quantized estimate of the cascaded channel, $\hat{\mathbf{g}}_{\mathrm{FB}}$, is available at the transmitter, the next step is to dynamically adjust the RIS phase configuration to maximize end-to-end system performance. This phase adaptation is particularly critical in time-varying environments, where the channel changes rapidly, and the RIS must track these variations in real time, despite practical limitations such as discrete phase resolution and limited feedback frequency.

To perform the adaptation, we employ a stochastic gradient–based control that updates the RIS phase vector based on the current channel estimate. The performance objective is
\begin{equation}
J(\boldsymbol{\phi}) \;=\; -\bigl|\hat{\mathbf{g}}^{H}\boldsymbol{\theta}\bigr|^{2}, 
\qquad \boldsymbol{\theta}=\bigl[e^{j\phi_{1}},\ldots,e^{j\phi_{N}}\bigr]^{T}.
\label{eq:obj_J}
\end{equation}
The $n$$^{th}$ phase is updated iteratively as
\begin{equation}
\phi_{n}^{(t+1)} \;=\; \mathcal{Q}_{b}\!\left[
\phi_{n}^{(t)} \;-\; \mu_{t}\,\mathrm{Im}\!\left\{ \hat{g}_{n}^{*}\,e^{j\phi_{n}^{(t)}}\,\hat{\mathbf{g}}^{H}\boldsymbol{\theta}^{(t)} \right\}
\right],
\label{eq:phase_update}
\end{equation}
where $\mu_{t}>0$ is the step size at iteration $t$. The operator $\mathcal{Q}_{b}[\cdot]$ denotes a $b$-bit \emph{uniform phase quantizer} that (i) wraps any input phase to $[0,2\pi)$ and (ii) snaps it to the nearest of $2^{b}$ equally spaced angles. Let $\Delta_b \!\triangleq\! 2\pi/2^{b}$ and $[\phi]_{2\pi}\!\triangleq\!\phi \bmod 2\pi$. Then
\begin{equation}
\mathcal{Q}_{b}(\phi)
\;=\;
\big(\,\Delta_{b}\,\mathrm{round}\!\big(\tfrac{[\phi]_{2\pi}}{\Delta_{b}}\big)\big)\bmod 2\pi,
\qquad
\Delta_{b}=\frac{2\pi}{2^{b}},
\label{eq:quantizer}
\end{equation}
which is equivalent to choosing the nearest codeword in
\begin{equation}
\mathcal{C}_{b}=\left\{\,0,\;\frac{2\pi}{2^{b}},\;\ldots,\;\frac{2\pi(2^{b}-1)}{2^{b}}\,\right\}.
\label{eq:quantizer_C}
\end{equation}

Intuitively, $\mathcal{Q}_{b}$ “rounds” any phase to the closest grid point; the pointwise error is bounded by $|\mathcal{Q}_{b}(\phi)-[\phi]_{2\pi}|\le \Delta_{b}/2$, and $\mathcal{Q}_{b}$ approaches the continuous-phase update as $b\to\infty$.

\noindent
For clarity, we now summarize \eqref{eq:obj_J}–\eqref{eq:quantizer} as a compact iteration that incorporates step-size scheduling, phase wrapping, and quantization projection.

\begin{algorithm}[t]
\caption{Adaptive RIS Phase Update with $b$-bit Quantization Projection}
\label{alg:adaptive_ris_phase}
\SetKwInOut{Input}{Input}\SetKwInOut{Output}{Output}
\Input{Channel estimate $\hat{\mathbf{g}}\!\in\!\mathbb{C}^{N}$; bits $b$; max iterations $T$; step sizes $\{\mu_t\}_{t=0}^{T-1}$; tolerance $\varepsilon$.}
\Output{Quantized phases $\boldsymbol{\phi}^{\star}\!\in[0,2\pi)^{N}$ and $\boldsymbol{\theta}^{\star}\!=\exp(j\boldsymbol{\phi}^{\star})$.}
\BlankLine
\textbf{Initialisation:} $\phi_n^{(0)} \leftarrow \arg(\hat{g}_n)$ (or any feasible); $\boldsymbol{\theta}^{(0)} \leftarrow \exp(j\boldsymbol{\phi}^{(0)})$.\;
\For{$t \leftarrow 0$ \KwTo $T-1$}{
  $s^{(t)} \leftarrow \hat{\mathbf{g}}^{H}\boldsymbol{\theta}^{(t)}$ \tcp*[r]{coherent sum}
  \For{$n \leftarrow 1$ \KwTo $N$}{
    $g^{\text{grad}}_n \leftarrow \mathrm{Im}\!\big\{ \hat{g}_n^{*}\, e^{j\phi_n^{(t)}}\, s^{(t)} \big\}$ \tcp*[r]{component of $\nabla_{\phi}J$}
    $\tilde{\phi}_n \leftarrow \phi_n^{(t)} - \mu_t\, g^{\text{grad}}_n$ \tcp*[r]{gradient step}
    $\phi_n^{(t+1)} \leftarrow \mathcal{Q}_b\!\big(\mathrm{mod}(\tilde{\phi}_n,2\pi)\big)$ \tcp*[r]{wrap to $[0,2\pi)$ then quantize}
  }
  $\boldsymbol{\theta}^{(t+1)} \leftarrow \exp\!\big(j\boldsymbol{\phi}^{(t+1)}\big)$\;
  \If{$\big|\hat{\mathbf{g}}^{H}\boldsymbol{\theta}^{(t+1)}\big| - \big|s^{(t)}\big| < \varepsilon$}{\textbf{break}}
  \tcp{Optional: update $\mu_{t+1}$ (constant, diminishing, or backtracking)}
}
$\boldsymbol{\phi}^{\star}\!\leftarrow\!\boldsymbol{\phi}^{(t)}$, \quad $\boldsymbol{\theta}^{\star}\!\leftarrow\!\boldsymbol{\theta}^{(t)}$.\;
\end{algorithm}

\noindent\textit{Remark:} 
The iteration incrementally aligns the RIS phases with the instantaneous channel while remaining compatible with finite-resolution optical hardware. Since optical phase shifters typically refresh at only a few kilohertz, the iteration rate must account for a millisecond-scale tuning latency, which, in turn, constrains the feasible pilot/feedback periodicity. For stability, the step size can follow standard gradient–descent conditions; in particular, with local curvature scaling as $\|\hat{\mathbf g}\|_2^{2}$, a sufficient range is $0 < \mu_t < 2/\|\hat{\mathbf g}\|_2^{2}$. In practice, a conservative choice (e.g., $\mu_t \le 1/\|\hat{\mathbf g}\|_2^{2}$ or a diminishing schedule) improves robustness against estimation noise. 

\textit{Compared to conventional RF RIS training,} the proposed adaptive scheme operates under fundamentally different constraints. In RF TDD systems, channel reciprocity enables the transmitter to infer the downlink from uplink pilots, allowing for rapid phase updates and dense pilot reuse. In contrast, optical wireless channels are generally non-reciprocal and feedback-limited; the downlink uses an optical carrier, whereas the uplink typically relies on a narrowband RF or infrared control link. Furthermore, optical receivers utilizing IM/DD architectures cannot directly measure phase, and turbulence-induced fading reduces the coherence interval. These factors motivate the use of processed, quantized feedback and a conservative, low-rate iterative update rather than full pilot retransmission or continuous analog control, as observed in RF RIS systems. The resulting algorithm reflects the inherent physical and architectural differences present in the optical domain.

With NMSE $\epsilon$ in the channel estimate, the forward-link SNR is well approximated by
\begin{equation}
\gamma_{\mathrm{eff}} \approx \gamma_f^\star(1-\epsilon).
\label{eq: Yeff}
\end{equation}
The corresponding capacity loss is bounded by
\begin{equation}
\Delta C \le \log_{2}\!\big(1+\gamma_f^\star\big)-\log_{2}\!\big(1+\gamma_f^\star(1-\epsilon)\big).
\label{eq:cap_loss}
\end{equation}
For small $\epsilon$, a first-order expansion of $\log_{2}(1+\gamma)$ at $\gamma_f^\star$ yields
\begin{equation}
\Delta C \approx \frac{\gamma_f^\star}{\ln 2\,\big(1+\gamma_f^\star\big)}\,\epsilon,
\label{eq:deltaC_firstorder}
\end{equation}
so $\Delta C$ grows approximately linearly with the NMSE. 

Expression~\eqref{eq:deltaC_firstorder} therefore establishes a clear, quantitative link between channel-estimation quality, system-design parameters (e.g., pilot and feedback budgets), and the resultant throughput degradation. It thus provides a practical design guideline for keeping the system within acceptable performance-loss margins under constrained feedback and estimation conditions.

\section{\label{sec: Feed }Complexity, Feedback, and Overhead Analysis}

The practical viability of adaptive RIS phase control depends not only on the accuracy of estimation and optimization but also on the complexity and overhead imposed by pilot signaling, CSI feedback, and control computation. This section presents a unified framework to quantify the resource requirements of the closed-loop architecture and delineates explicit design criteria for the system. To ground this analysis in practical settings, we consider representative parameter values from RIS-assisted OWC prototypes: the feedback spectral efficiency $\beta$ typically ranges from $0.5$ to $2$ bits/s/Hz, the feedback bandwidth $B_{\mathrm{FB}}$ spans from $50$ kHz to $2$ MHz depending on uplink technology, and the frame duration $T$ falls between $1$ and $10$ ms for real-time optical control or up to tens of milliseconds for quasi-static channels. These values are used to translate the feedback payload and update delay into realistic system-level constraints.

All CSI is conveyed from the transmitter to the receiver via a bandwidth-limited RIS-assisted uplink. The channel estimate $\hat{g}$ is an N-dimensional complex vector, with each component quantified using Q bits for the real part and Q bits for the imaginary part. Therefore, the total number of feedback bits required per coherence frame is:
\begin{equation}
B_{\mathrm{CSI}}=2\,N \,Q,\label{eq: Bcsi}
\end{equation}

To deliver these bits over the RIS-assisted uplink, the system allocates a fraction of each frame for feedback transmission. If the uplink supports a spectral efficiency of $\beta$ $[\mathrm{bit/s/Hz}]$ over a bandwidth of $B_{\mathrm{FB}}$ $[\mathrm{Hz}]$, the feedback capacity available in a frame of duration $T$ $[\mathrm{s}]$ is $\beta B_{\mathrm{FB}} T$ bits. Here, $T$ denotes the \emph{protocol scheduling frame} (pilot–estimation–feedback cycle), not the optical coherence time (typically microseconds under weak turbulence); thus, $T$ reflects the control-layer timescale for CSI refresh and RIS updates. The feedback can be conveyed via an auxiliary infrared control channel or a low-rate hybrid RF link, as noted in the introduction. The resulting feedback time fraction, which is the portion of the frame consumed by CSI feedback, is therefore:
\begin{equation}
\tau_{\mathrm{FB}}
=\frac{B_{\mathrm{CSI}}}{\beta B_{\mathrm{FB}} T}
=\frac{2 N Q}{\beta B_{\mathrm{FB}} T},
\label{eq:tauFB}
\end{equation}
where $B_{\mathrm{CSI}}$ is the per-frame payload (here modeled as $2NQ$ bits, i.e., two $Q$-bit quantities per RIS element).

This fraction directly quantifies the overhead incurred by feedback relative to the available transmission resources, establishing a clear link between quantization precision, the RIS array size, and the capacity of the feedback channel.

To ensure that sufficient time remains for payload data transmission, the
system enforces a minimum data duty cycle, denoted by $\eta_{\min}$.
This requires that the fraction of the frame allocated to feedback
does not exceed the allowable overhead; i.e., $\eta_{\min}$. the system must satisfy $\tau_{\mathrm{FB}}\leq1-\eta_{\min}$. Substituting
the earlier expression for $\tau_{\mathrm{FB}}$, we obtain the constraint
$\frac{2\:N\:Q}{\beta\:B_{\mathrm{FB}}\,T}\leq1-\eta_{\min}$. Rearranging for the quantization depth, the maximum allowable value of $Q$ is
\begin{equation}
Q\leq\frac{(1-\eta_{\min})\beta\, B_{\mathrm{FB}}\,\mathrm{T}}{2N}.
\end{equation}

This inequality directly links the RIS size $N$, feedback bandwidth $B_{\mathrm{FB}}$, frame duration $\mathrm{T}$, and maximum allowable quantization precision $\mathit{Q}$, ensuring that feedback transmission does not interfere with data throughput.

Additionally, the pilot overhead fraction is given by $\tau_{\mathrm{pilot}}=M/T.$ To ensure that the pilot and feedback signaling together do not exceed the available frame duration, their combined overhead must satisfy
\begin{equation}
\tau_{\mathrm{FB}}+\tau_{\mathrm{pilot}}\leq1-\eta_{\min},\Longleftrightarrow\quad\frac{M}{T}+\frac{2\,N\,Q}{\beta \,B_{\mathrm{FB}}\,T}\leq1-\eta_{\min}.\label{eq: TFB_2}
\end{equation}

This equation establishes a trade-off between pilot length $M$ and
quantization depth $Q$, both of which must be jointly optimized to
meet the system's throughput and feedback constraints. 
To achieve the target estimation accuracy (NMSE $\epsilon$), the pilot length must satisfy $M \ge \tfrac{N}{\epsilon\,\gamma_{\mathrm{pilot}}}$, where $\gamma_{\mathrm{pilot}}$ is the per-pilot SNR. Substituting this lower bound into the frame-overhead constraint yields the following \emph{feedback-quantization design bound} (maximum admissible quantization depth):
\begin{equation}
Q\leq\frac{\beta\, B_{\mathrm{FB}}}{2N}\left[(1-\eta_{\min})T-\frac{N}{\epsilon\:\gamma_{\mathrm{pilot}}}\right].\label{eq: Q}
\end{equation}

Inequality$\,$\eqref{eq: Q} specifies that, for a fixed $(N,\beta,B_{\mathrm{FB}},T,\eta_{\min},\gamma_{\mathrm{pilot}},\epsilon)$, the largest feasible $Q$ must meet the NMSE target while keeping the combined pilot and feedback overhead within the allowed data duty cycle.
Equation~\eqref{eq: Q} couples pilot SNR, target NMSE, quantization resolution (bits), feedback bandwidth, and RIS size \(N\), providing a sizing rule for training/feedback. Higher pilot SNR or a looser NMSE target permits fewer bits (smaller payload); lower SNR or a tighter NMSE requires more bits. Hence, pilot power, training length \(M\), and bit depth should be co-optimized to meet accuracy and latency within the feedback budget.

Now, considering computational complexity, the  LS with a generic \(\boldsymbol{\Phi}\) costs \(\mathcal{O}(MN+N^{2}+N^{3})\) per frame (the \(N^{3}\) term from inversion). With unitary \(\boldsymbol{\Phi}\) (\(\boldsymbol{\Phi}^{H}\boldsymbol{\Phi}=M\mathbf{I}\)), inversion is avoided, and the cost decreases to \(\mathcal{O}(MN)\). Quantizing \(\hat{\mathbf g}\) is \(\mathcal{O}(N)\), and each RIS phase update occurs \(\mathcal{O}(N)\) per iteration. Thus, optimal system performance is attained by jointly selecting the smallest feasible pilot length $M$ and the largest admissible quantization depth $Q$ such that the following two conditions are simultaneously satisfied: the channel estimation accuracy constraint,
\begin{equation}
\mathrm{NMSE}\leq\epsilon,
\end{equation}
and the total overhead constraint 
\begin{equation}
\tau_{\mathrm{pilot}}+\tau_{\mathrm{FB}}\leq1-\eta_{\min},
\end{equation}
Together, these criteria define the feasible design space for resource allocation and provide a principled foundation for the joint optimization of training and feedback in RIS-assisted optical wireless communication systems.

 \noindent\textit{Practical complexity and scalability,}
in LS implementations requiring matrix inversion, the computational load grows cubically with the number of RIS elements $N$; for $N\!\approx\!512$, the inversion cost reaches roughly $10^{9}$ real operations per frame, which can saturate a millisecond-scale processing budget. With unitary or orthogonal pilots, this inversion is avoided, and the cost reduces to $\mathcal{O}(MN)$, for example, $M\!=\!2N$ yields only $2N^{2}$ ($\approx5\times10^{5}$ mult-adds at $N\!=\!512$), which is three orders of magnitude lower. The LS updates can also be parallelized across RIS sub-arrays or GPU threads, making real-time adaptation feasible even for large arrays. Figures \ref{fig: 6}  and \ref{fig: 7} confirm this scaling trend and validate the derived feedback–NMSE trade-off, demonstrating that the proposed framework remains computationally practical for RIS sizes of up to a few-thousand-elements.

\section{\label{sec:Simu}Evaluation and Discussion of Results}
To assess the impact of phase quantization on channel-estimation accuracy, Monte Carlo simulations were conducted using the cascaded optical model described in Section \ref{sec:Sys_mod}. The model incorporates free-space path loss, Beer\textendash Lambert extinction, pixel-level diffraction, and independent turbulence on the $\mathrm{T\!\rightarrow\!RIS}$ and $\mathrm{RIS\rightarrow R}$ hops (cf. \eqref{eq: h=000026h}, \eqref{eq: Gpix}). The baseline geometry places the RIS on the plane $z=d_{TR}$ with the transmitter at the origin and the receiver positioned on boresight above the RIS $(x_{R}=y_{R}=0)$, ensuring that the diffraction variation across the aperture is primarily driven  by the elements' off-axis angles (cf. \eqref{eq: dis}).

\begin{table}[t]
  \centering
  \caption{Simulation parameters used for the quantization study.}
  \label{tab:quant_params}
  \small
  \resizebox{\columnwidth}{!}{%
  \begin{tabular}{@{}l c l@{}}
    \hline
    Quantity & Symbol & Value \\ \hline
    Optical wavelength & $\lambda$ & 1550\,nm \\
    Extinction coefficient & $\alpha$ & $10^{-4}\,\mathrm{m}^{-1}$ \\
    RIS lattice pitch (layout) & $p$ & 2\,cm \\
    Pixel size & $\Delta x=\Delta y$ & 2\,mm \\
    Tx--RIS range & $d_{\mathrm{TR}}$ & 1000\,m \\
    RIS--Rx height (boresight) & $z_R$ & 2500\,m \\
    Optical efficiency & $\eta_{\mathrm{opt}}$ & 0.7 \\
    RIS size & $N$ & $8\times 8$ ($N=64$) \\
    Pilot matrix & $\boldsymbol{\Phi}$ & Unitary DFT \\
    Pilot length & $M$ & $2N$ (unless stated) \\
    Pilot SNR & $\gamma_{\mathrm{pilot}}$ & 20\,dB (unless stated) \\
    Phase quantization & $Q$ & 1--8 bits \\
    Monte Carlo trials & --- & 200 per point \\
    \hline
  \end{tabular}}
\end{table}

For each realization, the cascaded channel vector $g$ was generated according to \eqref{eq: gn} using the parameters in Table~\ref{tab:quant_params}. LS estimation employed a DFT-based unitary pilot matrix (such that $\mathrm{\Phi^{H}\Phi=\mathit{M}I}$), consistent with the training model underpinning the CRLB/NMSE expression \eqref{eq: NMSE} and the pilot-length rule \eqref{eq:M_req}. Phase quantization was then emulated by uniformly quantizing only the estimated phase of each coefficient $g^{n}$ to $Q$ bits while preserving its estimated magnitude. We swept $\mathrm{Q\in[1,8]}$.

To benchmark the proposed jitter-aware, quantized-feedback framework, we include two idealized baselines that are commonly used in RIS-OWC studies.  \emph{(i) Ideal unquantized feedback:} perfect CSI and continuous RIS phase control (i.e., $M\!\to\!\infty$ or error-free least-squares (LS) with no delay; feedback resolution $b\!\to\!\infty$), which upper-bounds the achievable SNR/capacity and isolates degradation due to finite-rate feedback and imperfections in CSI.  \emph{(ii) Ideal diffraction-limited aperture without pointing jitter:} the Fraunhofer pattern with perfect alignment (i.e., $\Sigma^{(h)}\!=\!\mathbf{0}$, $S^{(h)}\!=\!1$, and the textbook $\mathrm{sinc}^{2}$ response of Sec.~II), which quantifies the power/gain shortfall attributable to long-exposure jitter averaging in practical optical channels.  Comparisons against these baselines make the relative gains of the proposed method transparent under realistic hardware and non-reciprocal feedback constraints.  In all reported results (unless otherwise noted), each data point is averaged over \emph{200} independent Monte Carlo trials; increasing the trial count beyond 200 alters NMSE, SNR, and capacity metrics by less than $1\%$, indicating statistical stabilization at this sampling depth.
\begin{figure}
\begin{centering}
\includegraphics[bb=0bp 0bp 484bp 291bp,width=4.1in,totalheight=2.5in,viewport=2bp 0bp 550bp 350bp]{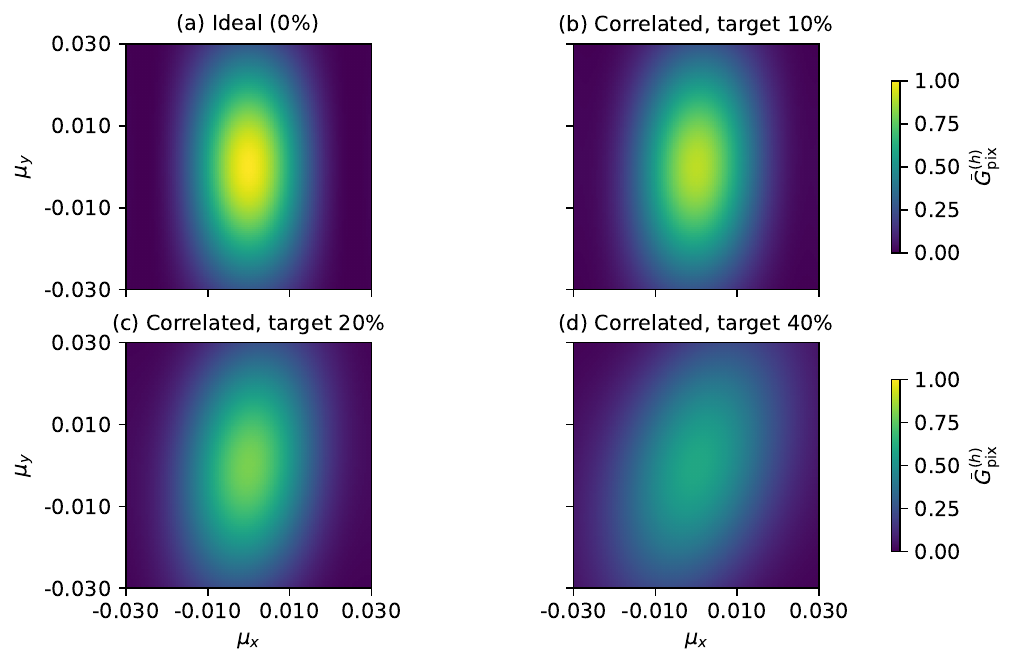}
\par\end{centering}
\caption{Comparison between ideal and Long-Exposure RIS pixel response under
angular jitter \label{fig: 1b }}
\end{figure}

Fig. \ref{fig: 1b } contrasts the ideal pixel response of Eq. \eqref{eq: Gpix} with the long-exposure response obtained by averaging over angular jitter according to Eq.\eqref{eq:LE_def}-\eqref{eq:perhop_integral}.
Subfigure (a) is the reference $\mathrm{sinc^{2}}$pattern from Eq.
\eqref{eq: Gpix}: a unit peak at boresight $(\mu_{x},\mu_{y})=(0,0)$
and deep nulls away from the axis, consistent with perfectly frontal,
jitter-free illumination. Subfigures (b)-(d) apply the correlated-jitter
model of Eq.\eqref{eq:LE_def}-\eqref{eq:perhop_integral} at settings
chosen to yield approximately 10\%, 20\%, and 40\%, boresight attenuation,
respectively. As the jitter variance increases, two linked effects
predicted by Eq.\eqref{eq:LE_def}-\eqref{eq:perhop_integral} become apparent: the boresight peak is reduced (long-exposure attenuation), and the sidelobe nulls are partially filled as energy spreads over nearby directions. These changes reflect the statistical averaging in Eq.\eqref{eq:LE_def}
and its closed-form Fourier/cosine representation in Eq. \eqref{eq:perhop_integral},
which together replace the unrealistic assumption of perfect alignment
with a physically accurate angular blur determined by the jitter covariance.

\begin{figure}[!t]
\centering
\includegraphics[width=4in, viewport=2bp 0bp 450bp 350bp]{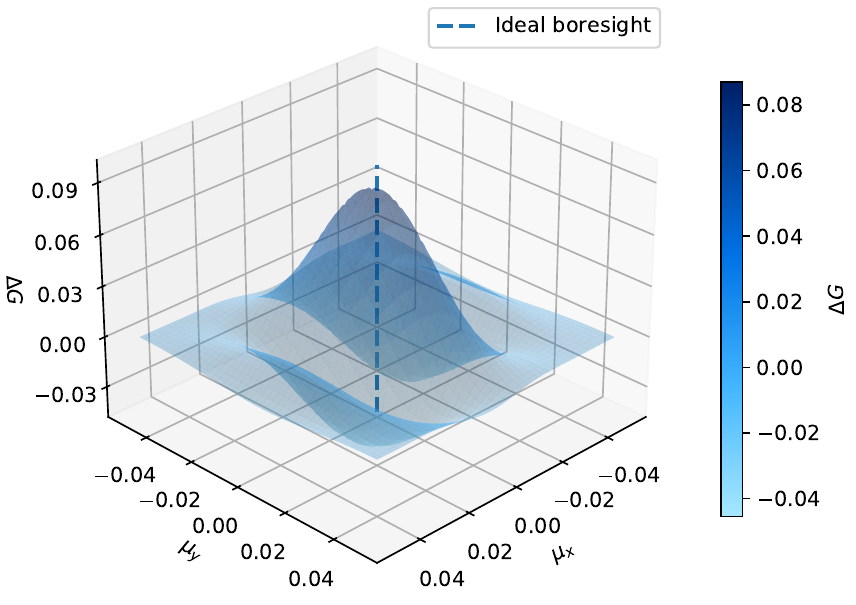}
\caption{Spatial distribution of the pixel-level gain deviation, $\Delta G = G_0 - \bar{G}_{\mathrm{pix}}$, between the ideal and long-exposure models under correlated angular jitter.}
\label{fig:2b2}
\end{figure}

To highlight the practical impact of jitter, Fig.~\ref{fig:2b2} illustrates the deviation surface $\Delta G(\mu_x,\mu_y)\!=\!G_0(\mu_x,\mu_y)-\bar{G}_{\text{pix}}^{(h)}(\mu_x,\mu_y)$, where $G_0$ is the ideal diffraction response and $\bar{G}_{\text{pix}}^{(h)}$ is the long-exposure per-hop response under the correlated-jitter covariance $\Sigma^{(h)}$ defined in~\eqref{eq:jitter_model}. Here, \(G_{0}\) represents the ideal diffraction response from Eq. \eqref{eq: Gpix}, and \(\bar{G}^{(h)}_{\text{pix}}\) denotes the long-exposure per-hop response from \eqref{eq:perhop_integral} under the correlated-jitter covariance \(\Sigma^{(h)}\) in \eqref{eq:jitter_model}. The surface measures, direction by direction, how the ideal pattern departs from the jitter-averaged pattern. Positive deviations concentrate near boresight \((\mu_x,\mu_y)\!\approx\!(0,0)\), where the ideal model predicts a unit peak that is reduced by pointing jitter; negative deviations appear around ideal nulls, reflecting energy that is redistributed off-axis. In short, jitter \emph{broadens} the main lobe, \emph{lowers} the peak, and \emph{fills} the nulls.
Because the long-exposure pixel gain in \eqref{eq:perhop_integral} feeds into the two-hop response in \eqref{eq:twohop}, these deviations accumulate across elements and directly affect the end-to-end link budget. The long-exposure model, therefore, avoids the boresight overestimation inherent to the ideal pattern and yields reliable performance predictions under realistic (imperfect) alignment, especially for large apertures where small per-pixel biases compound.

\begin{figure}
\begin{centering}
\includegraphics[width=3.5in,viewport=2bp 0bp 550bp 350bp]{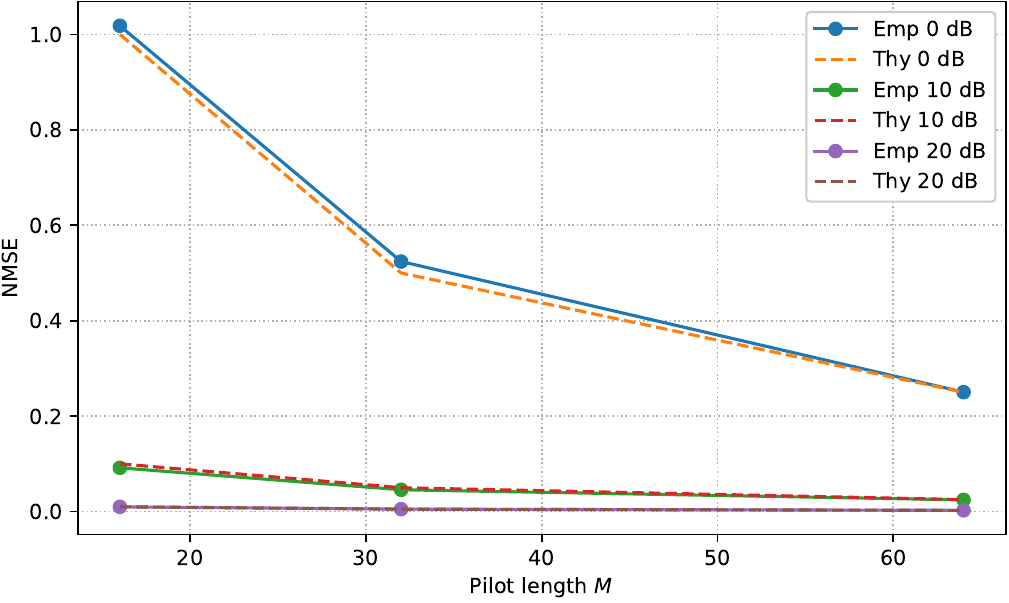}
\par\end{centering}
\caption{NMSE of the cascaded channel estimate vs. pilot length MM (unitary
training, LS estimator). \label{fig: 2b }}
\end{figure}

Fig. \ref{fig: 2b } presents the normalized NMSE performance of the proposed LS channel estimation scheme for a RIS-assisted OWC link, plotted as a function of the pilot length $M$ for three representative per-pilot SNR values of 0, 10, and 20\,dB. The solid curves correspond to Monte Carlo simulations based on a physically realistic cascaded channel model, while the dashed curves represent the analytical prediction in~\eqref{eq: NMSE}. The excellent overlap between the simulated and theoretical results across all SNR levels confirms that the LS estimator with a unitary (DFT-based) training matrix achieves the  CRLB for unbiased estimation under the Gaussian noise model. Because the CRLB defines the minimum achievable variance for any unbiased estimator, this agreement demonstrates that the proposed scheme is statistically efficient and fully exploits the available pilot information. Furthermore, the observed $1/M$ and $1/\gamma_{\mathrm{pilot}}$ scaling behaviors match the theoretical expectations, with longer pilot sequences and higher SNRs yielding proportionally lower estimation errors. This validates the analytical NMSE expression in~\eqref{eq: NMSE} and supports its use for pilot-length optimization and adaptive training design.  

  The NMSE–$M$ trend in Fig. \ref{fig: 2b }follows the analytical training-length rule derived in~\eqref{eq:M_req}, exhibiting the predicted $\sim 1/M$ decay at fixed $\gamma_{\mathrm{pilot}}$ and $N$. This consistency reinforces the reliability of the closed-form NMSE model for guiding real-time adaptation in practical RIS-assisted OWC systems.

\begin{figure}
\begin{centering}
\includegraphics[bb=0bp 0bp 484bp 291bp,width=3.5in,viewport=2bp 0bp 550bp 350bp]{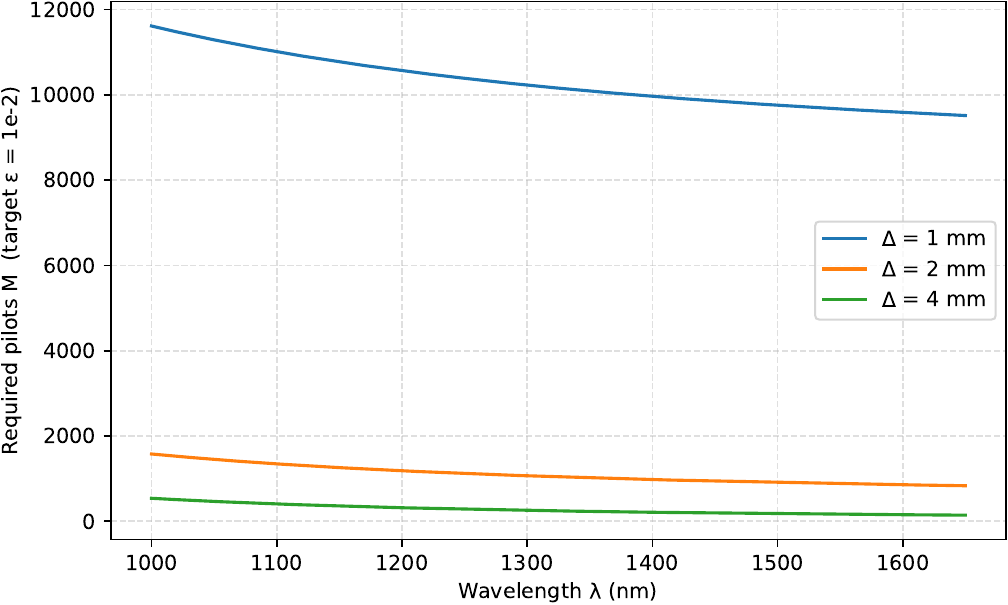}
\par\end{centering}
\caption{Required pilot length, $M$, versus wavelength and pixel size \label{fig: 1_2}}
\end{figure}

Fig. \ref{fig: 1_2} illustrates the required pilot length obtained  from the training length bound in \eqref{eq: NMSE 14}\textendash \eqref{eq:M_req}, where the normalized mean-square error is expressed as shown in \eqref{eq: NMSE}, and the minimum training length satisfies $M\geq N/M\:\gamma_{\mathrm{pilot}}$ with the structural floor $M\geq N$. The pilot SNR $\gamma_{\mathrm{pilot}}$ is computed from \eqref{eq: Ypilot} as the sum of the mean per-element cascaded powers, which incorporate pixel diffraction, Beer\textendash Lambert extinction, and free-space losses. To capture realistic long-exposure operation, the pixel response is averaged over a small pointing blur modeled as zero-mean Gaussian jitter, which introduces an explicit $\lambda$ dependence via $k=\tfrac{\pi\Delta}{\lambda}$ inside the
$\mathrm{sinc^{2}}(.)$factors. The resulting curves exhibit two consistent
trends. First, for any fixed $\Delta$, the required $M$ decreases
monotonically with $\lambda$ because the broader diffraction lobe
at longer wavelengths raises $\gamma_{\mathrm{pilot}}$ after long-exposure
averaging. Second, increasing $\Delta$ produces a pronounced reduction
in $M$ across the entire band, dominated by the $A^{2}$ scaling of the cascaded power; with sufficiently large pixels and at the long-wavelength end, the bound reaches the structural floor $M=N,$ implying that further increases in power or pixel size yield no additional reduction in training length. The absolute vertical placement of the curves depends linearly on $P_{T}/\sigma^{2}$ and inversely on the NMSE target $\varepsilon$; however, the monotonic ordering with respect to $\lambda$ and $\Delta$ is invariant and aligns with the theoretical predictions used to construct the bound.

\begin{figure}
\begin{centering}
\includegraphics[width=4in,totalheight=2.1in,viewport=2bp 0bp 550bp 350bp]{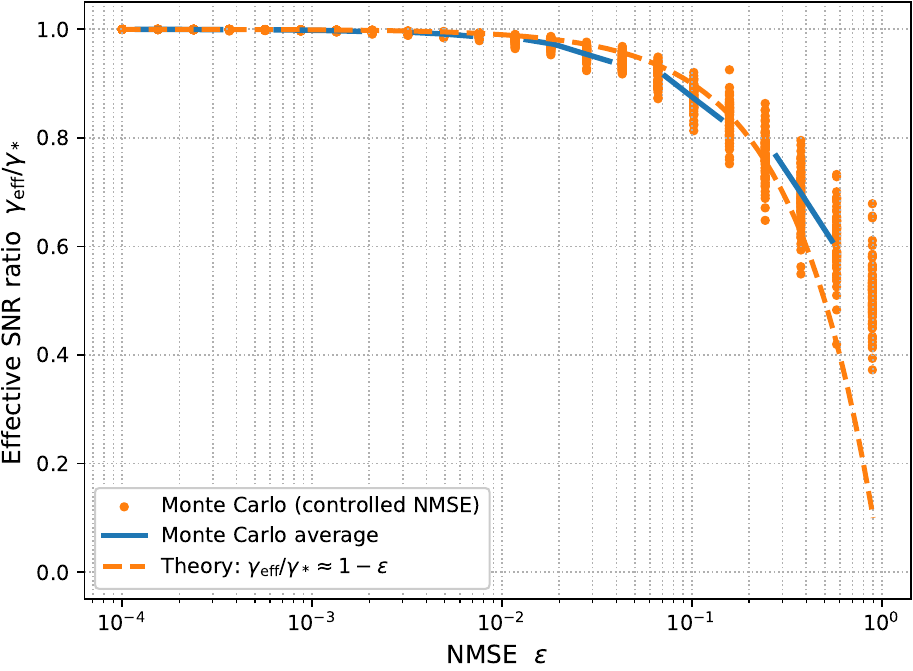}
\par\end{centering}
\caption{Effective SNR vs. channel-estimation NMSE. Points: Monte-Carlo; solid: bin-averaged trend; dashed: first-order model \(\gamma_{\mathrm{eff}}/\gamma_f^\star \approx 1-\varepsilon\) (Eq.~\eqref{eq: Yeff}). Close agreement for small \(\varepsilon\) validates the NMSE\(\rightarrow\)SNR conversion used for pilot/feedback sizing. \label{fig:4}}
\end{figure}

Fig.~\ref{fig:4} quantifies the forward-link SNR loss when the RIS phases are computed from an imperfect cascaded-channel estimate. each realization perturbs the true channel by an isotropic complex error scaled to a target NMSE $\varepsilon$ (cf.~\eqref{eq: NMSE}), and the resulting SNR, using the estimated phases, is normalized to the perfect-CSI benchmark. The Monte-Carlo samples and their bin-averaged trend closely match the analytical prediction in \eqref{eq: Yeff}, establishing a near-linear SNR penalty versus $\varepsilon$ in the small-error regime. Consequently, once an allowable SNR degradation is specified, the corresponding NMSE target can be read from the curve and translated, via \eqref{eq: Ypilot} and \eqref{eq:M_req}, into concrete pilot-SNR, pilot-length, and feedback-resolution requirements.

\begin{figure}[t]
\centering
\includegraphics[width=3.7in,viewport=2bp 0bp 550bp 350bp]{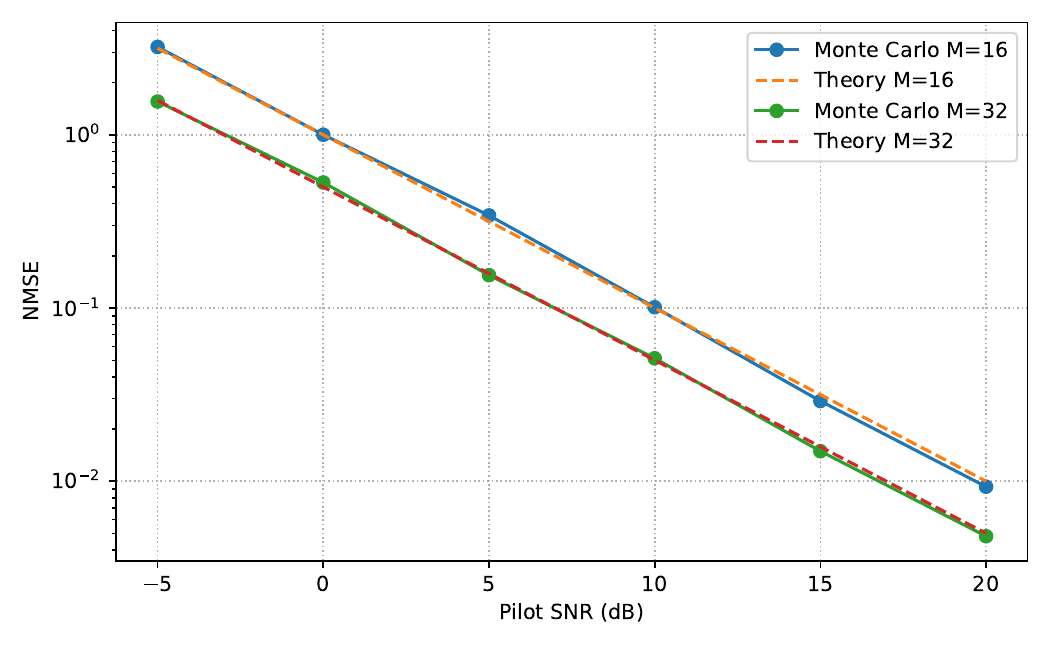}
\caption{Effect of pilot SNR and pilot length on cascaded-channel NMSE (unitary training, LS). 
Solid curves: Monte Carlo simulations; dashed curves: Theoretical prediction 
($\mathrm{NMSE}=N/(M\,\gamma_{\mathrm{pilot}})$).}
\label{fig:3b}
\end{figure}

  Fig.~\ref{fig:3b} shows that the LS estimator’s NMSE follows the law $\mathrm{NMSE}\!\approx\!N/(M\,\gamma_{\mathrm{pilot}})$ under unitary training: on a logarithmic scale, each curve has a $-1$\,dB-per-dB slope versus pilot SNR, and doubling the pilot length ($M=2N$) shifts the curve down by $\approx3$\,dB. With $\Phi^{H}\Phi=M I$, the LS estimator attains the Gaussian CRLB, so the $(M,\gamma_{\mathrm{pilot}})$ pair needed for any target NMSE can be read directly from the closed-form expression. The small-error mapping $\gamma_{\mathrm{eff}}/\gamma_f^{\star}\!\approx\!1-\varepsilon$ converts NMSE to SNR loss (e.g., $\varepsilon=0.005\Rightarrow 0.5\%$), confirming the resource-allocation rules used throughout.

\begin{figure}
\begin{centering}
\includegraphics[bb=0bp 0bp 484bp 330bp,width=4in,totalheight=2.2in,viewport=2bp 0bp 550bp 350bp]{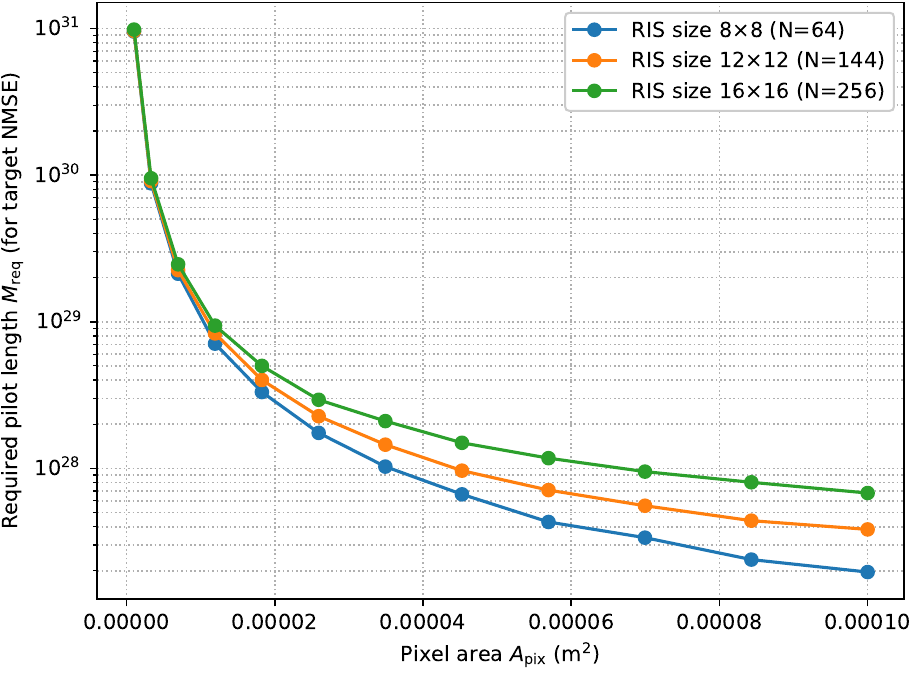}
\par\end{centering}
\caption{Impact of RIS pixel area on training overhead (boresight receiver)
\label{fig:2b}}
\end{figure}

Fig.~\ref{fig:2b} shows that the pilot length \(M\) required for a fixed estimation accuracy decreases with the RIS pixel area and increases with the array size \(N\). By \eqref{eq: E=gn_abs}, the per-element mean power \(\mathbb{E}[|g_n|^2]\) scales approximately as \(A_n^2\) (with diffusion and range factors), which raises the aggregate pilot SNR \(\gamma_{\mathrm{pilot}}\) in \eqref{eq: Ypilot}; via \eqref{eq: NMSE}, higher \(\gamma_{\mathrm{pilot}}\) lowers the LS NMSE and, through \eqref{eq:M_req}, reduces the required \(M\). Hence, small pixels yield weak \(\mathbb{E}[|g_n|^2]\), low \(\gamma_{\mathrm{pilot}}\), and large \(M\); whereas larger \(A_n\) strengthen the link and reduce \(M\). The vertical separation between curves with different \(N\) reflects the explicit proportionality \(M\!\propto\!N\) in \eqref{eq:M_req}; larger arrays require more pilots at the same target NMSE unless compensated by a higher \(\gamma_{\mathrm{pilot}}\). Overall, the figure shows that stronger per-pixel optics (larger $A_{n}$ or favorable geometry) reduce the required pilot length $M$, while a larger RIS (higher $N$) increases $M$, as predicted by
 \eqref{eq: E=gn_abs}-\eqref{eq: Ypilot}-\eqref{eq: NMSE}\textendash \eqref{eq:M_req}.

\begin{figure}
\begin{centering}
\includegraphics[width=4in,totalheight=2.3in,viewport=2bp 0bp 550bp 350bp]{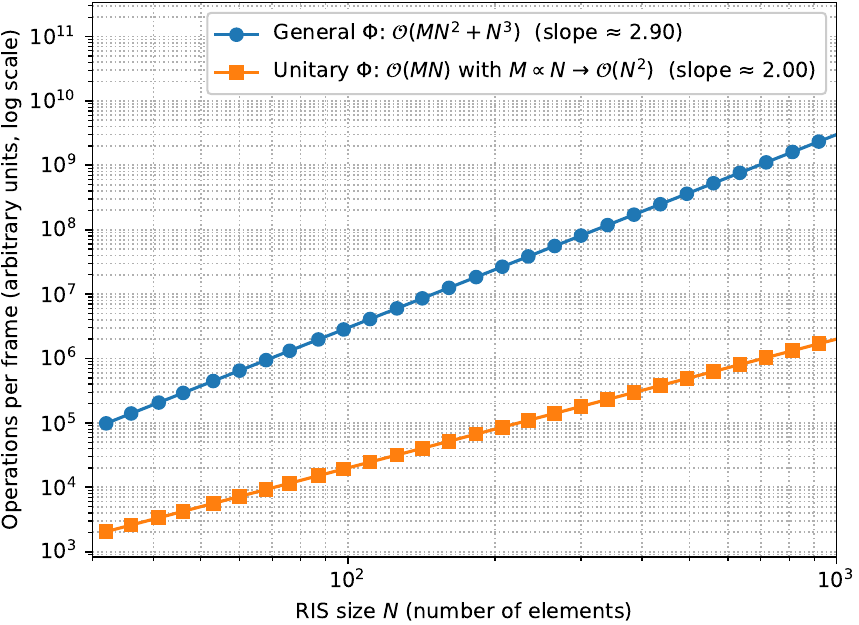}
\par\end{centering}
\caption{Computational demand of LS channel estimation versus RIS size\label{fig: 6}}
\end{figure}

Fig \ref{fig: 6}. reports the per\textendash coherence-frame operation
count required for LS channel estimation as the number of RIS elements $N$ increases, contrasting a general pilot $\Phi$ with a unitary (orthogonal) matrix $\Phi$. With a general $\Phi$, each frame must form the Gram matrix and invert an $N\times N$ system, leading to a complexity of order $\mathrm{\mathcal{O}(\text{M\,\ensuremath{N^{2}}+\ensuremath{N^{3}}})}$
(dominated by the cubic term), as stated in the complexity discussion. In contrast, when a unitary pilot structure is used such that $\Phi H\Phi=MI$
, the matrix inversion is avoided, and the LS computation reduces to
a single matrix\textendash vector product of order $\mathcal{O}(M\:N)$
per frame. Because the training length $M$ must scale linearly with $N$ to meet identifiability and accuracy targets (see the pilot-length rule and its unitary-pilot specialization), the practical asymptotics
become $\mathcal{O}(N^{3})$ for general $\Phi$ and $\mathcal{O}(N^{2})$
for unitary $\Phi$. The widening gap in the plotted curves at large
$N$ therefore confirms the theoretical recommendation to employ unitary
training for scalable RIS deployments, removing the inversion bottleneck
while preserving LS optimality under the model.
\begin{figure}
\begin{centering}
\includegraphics[width=4in,totalheight=2.2in,viewport=2bp 0bp 550bp 350bp]{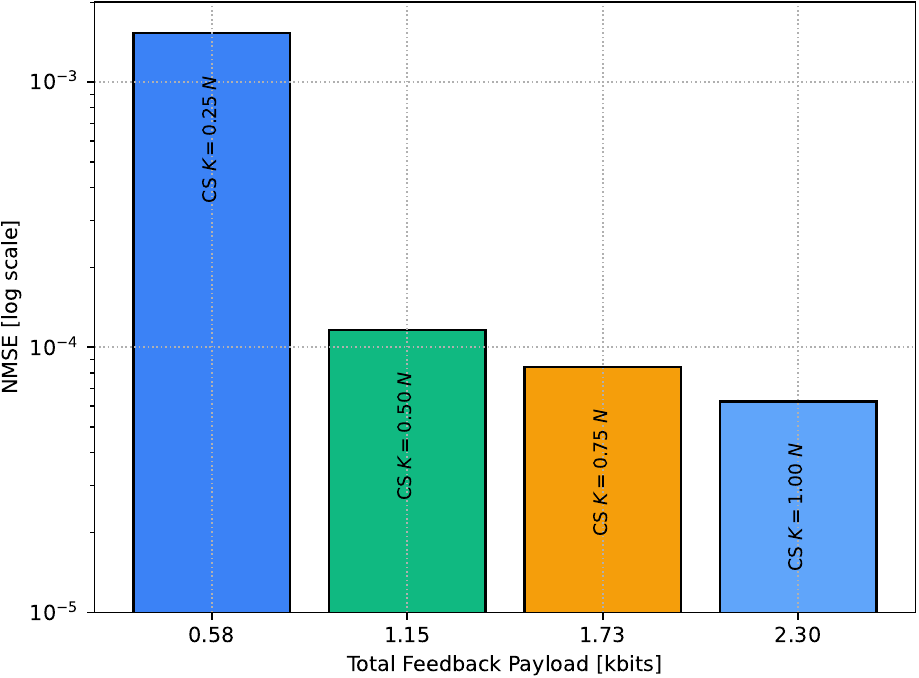}
\par\end{centering}
\caption{NMSE vs. total feedback payload using compressed-sensing (CS) feedback. Bars correspond to measurement ratios \(K/N\in\{0.25,\,0.50,\,0.75,\,1.00\}\) (legend: “CS \(K=0.25N\)”, “CS \(K=0.50N\)”, “CS \(K=0.75N\)”, “CS \(K=1.00N\)”). The abscissa reports total feedback bits per coherence interval; in the CS setting this accounting replaces \(N\) by \(K\) in the payload expressions (i.e., only the \(K\) compressed measurements are fed back). As the payload increases (larger \(K/N\) or more bits), the NMSE drops sharply and then saturates at a common estimator/noise floor determined by training SNR and pilot length, consistent with the prediction in Eq.~(32) as \(K\!\to\!N\) and quantization resolution increases.}
\label{fig: 7}
\end{figure}

Figure  \ref{fig: 7} presents the empirical relationship between feedback payload and channel-estimation NMSE when the receiver employs compressed-sensing (CS) feedback in the RIS-assisted optical link. In the proposed pipeline, the receiver first forms a pilot-aided channel estimate and then exploits transform sparsity prior to feedback by projecting the estimate onto a sparsifying basis and encoding only \(K\) compressed measurements (cf. the processed/CS mapping). Each bar in Fig.~\ref{fig: 7} corresponds to a distinct compression ratio \(K/N\in\{0.25,0.50,0.75,1.00\}\), and its horizontal position equals the total number of feedback bits per coherence interval, using the same payload accounting as in the uncompressed case, but with \(N\) replaced by \(K\). As the payload increases, either by enlarging \(K\) (higher \(K/N\)) or by allocating more bits per measurement, the NMSE decreases by orders of magnitude and approaches a common floor. This floor reflects the estimator/noise limit set by the pilot SNR and training length and matches the analytical behavior in Eq.~(32): the CS-induced distortion vanishes as \(K\!\to\!N\) (and quantization resolution increases); thus, the residual error is governed by the uncompressed estimator. Interpreted within the system budget, the abscissa also maps to the fraction of frame time consumed by feedback via the timing expression for the uplink; hence, selecting \(K\) transparently trades estimation accuracy for feedback time. The combined pilot–feedback constraint enforces feasibility within the overall frame duration.
 
\section{\label{sec:Conclusion}Conclusion}

We introduced a physically consistent long-exposure optical RIS model integrated with an IM/DD--compatible unitary-pilot LS estimator and a quantization-aware processed-feedback path for non-reciprocal uplinks. The framework yields closed-form sizing rules that tie pilot length, pixel geometry, and feedback resolution to target NMSE and capacity loss while avoiding per-frame inversions and preserving near-quadratic complexity in array size. In a representative setting (\(N=64\), pilot SNR \(=20\,\mathrm{dB}\), \(M=2N\)), the NMSE is \(\approx 5\times 10^{-3}\), implying a \(\sim 0.5\%\) effective-SNR loss and a \(\approx 7\times 10^{-3}\,\mathrm{bits/s/Hz}\) capacity penalty; 6-bit phase feedback operates within \(\le 0.5\,\mathrm{dB}\) of the estimator floor. Training scales with pixel geometry (halving pixel width \(\Rightarrow \sim 4\times\) pilots), while atmospheric extinction bounds feasible feedback rates; enlarging the RIS aperture recovers several decibels of uplink margin. These prescriptions enable practical link budgeting and feedback-rate planning for large-aperture optical RIS links. Future directions include multi-RIS/multi-user extensions, non-Gaussian or temporally correlated jitter, rate-distortion-aware feedback scheduling, and hardware-in-the-loop trials.

\bibliographystyle{IEEEtran}
\bibliography{Optic}

\end{document}